\documentclass[12pt]{article}

\usepackage{amssymb}
\usepackage{subfig}
\usepackage{color}
\usepackage{epsfig,amssymb,amsfonts,amsmath,graphicx,dsfont,cite,xfrac}
\usepackage{authblk}
\definecolor{mygray}{gray}{0.5}

\usepackage{cite}
\usepackage[colorlinks=true,linkcolor=blue,citecolor=red]{hyperref}

\newcommand{\be}{\begin{equation}}
\newcommand{\ee}{\end{equation}}
\newcommand{\bea}{\begin{eqnarray}}
\newcommand{\eea}{\end{eqnarray}}

\title{Constructing Squeezed States of Light with Associated Hermite Polynomials}

\author[${1,3}$]{K. Zelaya\thanks{Corresponding author; E-mail:  zelayame@crm.umontreal.ca}}
\author[${1,2}$]{V. Hussin\thanks{veronique.hussin@umontreal.ca}}
\author[${3}$]{O. Rosas-Ortiz\thanks{orosas@fis.cinvestav.mx}}

\affil[${1}$]{\footnotesize Centre de Recherches Math\'ematiques, Universit\'e de Montr\'eal, Montr\'eal H3C 3J7, QC, Canada}

\affil[${2}$]{\footnotesize D\'epartement de Math\'ematiques et de Statistique, Universit\'e de Montr\'eal, Montr\'eal H3C 3J7, QC, Canada}

\affil[${3}$]{\footnotesize Physics Department, Cinvestav, AP 14-740, 07000
M\'exico City, Mexico}

\date{}
\begin{document}

\maketitle

\begin{abstract}

A new class of states of light is introduced that is complementary to the well-known squeezed states. The construction is based on the general solution of the three-term recurrence relation that arises from the saturation of the Schr\"odinger inequality for the quadratures of a single-mode quantized electromagnetic field. The new squeezed states are found to be linear superpositions of the photon-number states whose coefficients are determined by the associated Hermite polynomials. These results do not seem to have been noticed before in the literature. As an example, the new class of squeezed states includes superpositions characterized by odd-photon number states only, so they represent the counterpart of the prototypical squeezed-vacuum state which consists entirely of even-photon number states.

\end{abstract}


\section{Introduction}

The uncertainty principle represents one of the most distinctive features of quantum mechanics. In contrast to classical theories, this principle denies the possibility of having exact values for simultaneous measurements of canonically conjugated physical variables. Discovered by Heisenberg in 1927 for the position and momentum of an electron ``observed'' under an idealized microscope \cite{Hei27}, this quantum law has undergone improvements while intense discussion has been developed about its interpretation and meaning \cite{Hil16,Jam66,Grupo1}. Heisenberg's discovery implies that exact knowledge of the electron position produces a wide spread of its momentum, ``the more precisely the position is determined, the less precisely the momentum is known'' \cite{Hei27}, and vice versa. Although the initial glimpse of this result may be traced back to Dirac and Jordan \cite{Jam66}, it was Heisenberg who opened the general question about what can be and what cannot be measured in quantum theory \cite{Mie09}. After some refinements made independently by Kennard \cite{Ken27}, Condon \cite{Con29}, and Robertson \cite{Rob29}, a more precise mathematical formulation was obtained by Schr\"odinger in terms of the Schwartz inequality \cite{Sch30}.

The correct determination of the minimum values associated to the uncertainty principle makes sense because the quantum formulation is strongly linked to experimental measurements: the observable quantities define what can be included into the theory (Heisenberg) and, at the same time, they are decided by the theory itself (Einstein). 

On the other hand, the trend of looking for minimum uncertainty states embraces a large number of relevant works throughout different stages of modern quantum theory  \cite{Ros19}. Pioneering results include the Schr\"odinger's wave packets of constant width \cite{Sch26} (representing a mathematical  antecedent of the {\em coherent states} introduced by Glauber in quantum optics \cite{Gla07}), followed almost immediately by the Kennard's wave packets of oscillating time-dependent width \cite{Ken27} (representing the first example of what are nowadays called {\em squeezed states} \cite{Hol79}). 

The celebrated Glauber states \cite{Gla07} started the practical applications of quantum theory in optics by extending the notion of optical coherence to the description of quantized electromagnetic fields \cite{Ros19}. According to Glauber, the fully coherent states of the quantized electromagnetic radiation tolerate a description in terms of the Maxwell theory, so they may be considered `classical'. These fascinating quantum states are eigenvectors of the boson-annihilation operator with complex eigenvalue \cite{Gla07}, and minimize both the Heisenberg-Kennard and Schr\"odinger inequalities\footnote{Heisenberg \cite{Hei27}, Kennard \cite{Ken27}, Condon \cite{Con29} and Robertson \cite{Rob29} taken into account only two of the three quadratic moments that can be associated with two variables. Schr\"odinger \cite{Sch30} was the first to notice that the covariance $\sigma_{A,B}$  must be considered together with the variances  $(\Delta A)^2$ and $(\Delta B)^2$ in order to better define the lower bound of the uncertainty principle for $A$ and $B$. Nevertheless, it is a common mistake to quote the main result of the Schr\"odinger paper \cite{Sch30} as the Schr\"odinger-Robertson inequality. Throughout this work we opt by qoting $(\Delta A)^2 (\Delta B)^2 \geq  \sigma_{A,B}^2 + \tfrac14 \vert \langle [A,B] \rangle \vert^2$ as the {\em Schr\"odinger inequality}.}. In position-representation the Glauber states acquire the form of the Schr\"odinger wave packets, a fact that is usually interpreted in favor of Schr\"odinger as the precursor of coherent states (see however  the discussion about the different nature of the Glauber and Schr\"odinger results in \cite{Ros19}). 

An additional class of states of light is such that one of the variances is squeezed at the time that the other is stretched in order to minimize the uncertainty inequality of the field quadratures \cite{GrupoS}. These states are called squeezed (a term coined in the field of gravitational-wave detection \cite{Hol79}, see also \cite{Sch10,Bar19}), and find immediate applications in interferometry \cite{Har03}. In contrast to the Glauber states, the squeezed states of light admit no description in terms of the Maxwell theory, so they are {\em nonclassical} \cite{Ros19} (a very comprehensive account of the evolution in the study of nonclassical states can be found in the collection of reviews edited by Dodonov and Man'ko \cite{Dod03}).

In this work we introduce a new class of squeezed states that can be expressed as a linear superposition of photon-number states where the vacuum state is absent. Such a class is complementary to the already known family of squeezed states that includes the squeezed-vacuum as a distinguished example. In fact, while the squeezed-vacuum consists entirely of even-photon number states, the new class includes a counterpart consisting on odd-photon number states only. The striking profile of the number-state superpositions giving rise to the new squeezed states is that the related probability amplitudes are characterized by the associated Hermite polynomials \cite{Ask84} (see also \cite{Mou05}). Such polynomials satisfy a three-term recurrence relation quite similar to that fulfilled by the classical Hermite polynomials but presenting a slight alteration in the labeling of the recurrence coefficients, which modifies the corresponding initial values. The new class of squeezed states is therefore defined by a set of associated Hermite polynomials parameterized by two complex numbers, which also characterize the nonclassical properties of such states.

In Section~\ref{section2} we revisit some generalities of the minimum uncertainty states in order to obtain the three-term recurrence relation we are interested in.  Then, a  general approach is proposed where the recurrence relations are treated as finite difference equations, the general solution of which is obtained before considering the initial conditions. The squeezed-vacuum states are recovered as immediate example while a new class of squeezed states consisting entirely of odd-photon number states is introduced. Section~\ref{section3} is devoted to the construction of squeezed states in terms of the associated Hermite polynomials, hereafter referred to as {\em associated squeezed states}. The conventional squeezed states, expressed in terms of the classical Hermite polynomials, are also recovered to show the applicability of the method. Section~\ref{section4} is addressed to the study and discussion of the nonclassical properties of the associated squeezed states. In Section~\ref{conclu} the main results of the work are remarked. For self-consistency, we have included two appendices with detailed information concerning the main points of our approach as well as the generalities of the orthogonal and associated polynomials.

\section{Single-mode minimum uncertainty states}
\label{section2}

The Schr\"odinger inequality 
\be
(\Delta A)^2 (\Delta B)^2 \geq  \sigma_{A,B}^2 + \tfrac14 \vert \langle C \rangle \vert^2, \quad \sigma_{A,B} = \tfrac12 \langle A B + B A \rangle -\langle A \rangle \langle B  \rangle,
\label{AB}
\ee
assumes that the expectation values are evaluated with regular ({\em normalizable}) vectors in the Hilbert space $\mathcal{H}$ where the observables $A$ and $B$ are represented by self-adjoint operators. For  $A= B$ one has $\sigma_{A,A} = \langle A^2 \rangle - \langle A \rangle^2 = (\Delta A)^2$, so the variance is a particular case of covariance. Besides, whereas the expectation value $\langle C \rangle$ is attainable to the commutation properties of observables $A$ and $B$, the covariance $\sigma_{A,B}^2$ is mainly determined by the statistical independence of such variables. Thus, the Schr\"odinger inequality takes into account both the statistical properties of the observables to be measured and the states used in the measurement. 

We may identify at least three classes of quantum states: (i) States producing null covariance $\sigma_{A,B} =0$. These are such that the Condon-Robertson and Schr\"odinger inequalities are equivalent \cite{Pur94}, so the lower bound of the uncertainty principle is unambiguously established. The Glauber states are prototypical of this class for the field quadratures (ii) States for which neither $\sigma_{A,B}$ nor $\langle C \rangle$ are equal to zero. This class of states satisfies the uncertainty principle in terms of the Schr\"odinger inequality. Most of the squeezed states reported in the literature find place in this class for the field quadratures (iii) States for which $\langle C \rangle = 0$. By necessity, this class requires the Schr\"odinger inequality to define the lower bound of uncertainties because the Condon-Robertson inequality becomes redundant \cite{Pur94}.

We would like to emphasize that even the Schr\"odinger inequality (\ref{AB}) could fail in determining the lower bound of the uncertainties for the third group mentioned above. For if $\langle C \rangle=0$, then $\langle AB \rangle = \langle BA \rangle$, and $\sigma_{A,B} = \langle AB \rangle - \langle A \rangle \langle B \rangle$. If now $\sigma_{A,B} =0$, we immediately get $\langle A B \rangle = \langle A \rangle \langle B \rangle$, which makes redundant the inequality (\ref{AB}) when both $\Delta A$ and $\Delta B$ are different from zero \cite{Pur94}. Nevertheless, a form of saturating (\ref{AB}) with zeros at both sides would be to consider eigenstates of either $A$ or $B$. In such case the identity $\langle A B \rangle = \langle A \rangle \langle B \rangle$ is automatically satisfied and the variances of $A$ and $B$ are both cancelled. The latter means statistical independence between the observables $A$ and $B$. 

Avoiding the redundant cases, and paying attention to states for which $A$ and $B$ are not statistically independent, it may be shown that inequality (\ref{AB}) is saturated by the solutions of the eigenvalue equation \cite{Jac68,Sto69}
\be
(A - i \lambda B) \vert \beta \rangle = \beta \vert \beta \rangle, \quad \lambda \in \mathbb C, \quad \beta = \langle A \rangle -i \lambda \langle B \rangle.
\label{AB2}
\ee

\subsection{Eigenvalue equation in the Fock basis}

We are interested in solving Eq.~(\ref{AB2}) for $A=x = \hat x \sqrt{m\omega/\hbar}$ and $B= p =\hat p/ \sqrt{m\omega \hbar}$, with $x = \frac{1}{\sqrt 2} (a^{\dagger} + a)$ and $p = \frac{i}{\sqrt 2} (a^{\dagger} -a)$. Hereafter $a$ and $a^{\dagger}$ stand for the boson-ladder operators $[a, a^{\dagger}]=1$. From (\ref{AB2}) we have $x - i \lambda p = \frac{1}{\sqrt 2} [(1+ \lambda) a + (1-\lambda) a^{\dagger}]$, which may be rescaled as follows
\begin{equation}
(a+\xi a^{\dagger})\vert \alpha , \xi \rangle = \alpha \vert \alpha,\xi \rangle, \quad \alpha,\xi\in\mathbb{C},
\label{eq:NSS4}
\end{equation}
where $\xi =\frac{1+\lambda}{1-\lambda}$, $\alpha = \frac{\beta{\sqrt 2}}{1-\lambda}$, and $\lambda \neq 1$ \cite{Fu96,Alv02,Dey15,Zel18a,Dey18,Dey20}.

\begin{figure}[htb]
\centering
\includegraphics[width=0.4\textwidth]{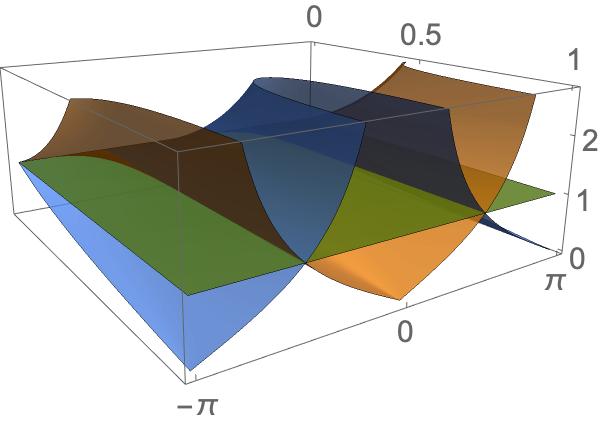}

\caption{\footnotesize
Squeezing parameter $\tau$ (orange surface) and its reciprocal $\tau^{-1}$ (blue surface) for $\xi=r e^{i\theta}$, with $r \in [0,1)$ and $\theta \in [-\pi, \pi)$. These parameters define the squeezing properties of the variances $(\Delta x)^2$ and $(\Delta p)^2$ calculated with the minimum uncertainty states $\vert \alpha, \xi \rangle$, see Eqs.~(\ref{eq:NSS4})-(\ref{schro}). The value $\tau=1$ (green surface) is included as a reference. 
}
\label{Fsqueezing}
\end{figure}

Assuming that we have at hand the regular solutions of (\ref{eq:NSS4}), the straightforward calculation shows that the variances of $x$ and $p$ can be written in the form
\begin{equation}
(\Delta x)^2 = \tau \sqrt{\tfrac{1}{4} + \sigma^{2}_{x,p}}, \qquad  (\Delta p)^2= \frac{1}{\tau} \sqrt{\tfrac{1}{4} + \sigma^{2}_{x,p}}, \qquad \tau = \left\vert \tfrac{1-\xi}{1+\xi} \right\vert.
\label{eq:NSS3-x}
\end{equation}
Thus, $\vert \alpha, \xi \rangle$ saturates the Schr\"odinger inequality for the field quadratures 
\be
(\Delta x)^2 (\Delta p)^2 \geq \tfrac14 + \sigma_{x,p}^2, \quad \sigma_{x,p} = \tfrac12 \langle xp +px \rangle - \langle x \rangle \langle p \rangle.
\label{schro}
\ee
Using $\xi=r e^{i\theta}$, with $r \in [0,1)$ and $\theta \in [-\pi, \pi)$, the $\tau$-parameter in (\ref{eq:NSS3-x}) can be classified as follows:
\begin{itemize}
\item[a)] 
$\tau=1$ for either $r=0$ or $\theta=\pm \tfrac12 \pi$ and any allowed value of $r$.
\item[b)] 
$0< \tau <1$ for  $r \in (0,1)$ and $\vert \theta \vert < \tfrac12 \pi$.
\item[c)] 
$\tau >1$ for $r \in (0,1)$ and $\tfrac12 \pi< \vert \theta \vert \leq \pi$.
\end{itemize}
Figure~\ref{Fsqueezing} shows the behavior of $\tau$ (orange surface) and $\tau^{-1}$ (blue surface) in terms of $r$ and $\theta$. The value $\tau=1$ (green surface) is included as a reference.

The solutions of the eigenvalue equation (\ref{eq:NSS4}) provide a repository of minimum uncertainty states $\vert \alpha, \xi \rangle$ that may produce the squeezing of either $(\Delta x)^2$ or $(\Delta p)^2$ for the appropriate squeezing parameter $\tau$. For instance, the case (b) implies that the variances $(\Delta x)^2$ and $(\Delta p)^2$ are respectively squeezed and stretched. They interchange roles for the parameters defined in case (c). The states leading to any of the above cases are called squeezed \cite{GrupoS,Hol79}. In turn, case (a) corresponds to the Glauber states for $\xi = r=0$ since  $(\Delta x)^2 = (\Delta p)^2 = \tfrac12$, with $\sigma_{x,p}^2=0$. On the other hand, fixing $\theta = \pm \frac12 \pi$ means $\xi = \pm i r$, so that $(\Delta x)^2 = (\Delta p)^2$, with $\sigma_{x,p}^2$ parameterized by $r$.

\subsection{Three-term recurrence relation for squeezing}
\label{3term}

The construction of regular vectors $\vert \alpha, \xi \rangle \in \mathcal{H}$ fulfilling the eigenvalue equation (\ref{eq:NSS4}) is usually addressed by writing them in terms of the  Fock basis $\vert n \rangle$, $n=0,1,2,\ldots$ Namely,
\begin{equation}
\vert \alpha,\xi \rangle = \sum_{n=0}^{\infty} \mathcal{C}_{n}(\alpha,\xi) \vert n \rangle, 
\label{eq:NSS4-1}
\end{equation}
with
\begin{equation}
\mathcal{C}_{n}(\alpha,\xi)=\frac{1}{\mathcal{N}(\alpha,\xi)} \frac{P_{n}(\alpha,\xi)}{\sqrt{n!}}, \quad n=0,1,2, \ldots
\label{eq:NSS4-2}
\end{equation}
The normalization factor $\mathcal{N}(\alpha,\xi)$ may be defined by demanding the convergence of the series
\be
\left\vert \mathcal{N}(\alpha,\xi) \right\vert^2 = \sum_{n=0}^{\infty} \frac{\vert P_n(\alpha, \xi) \vert^2}{n!}.
\label{factor}
\ee
Introducing \eqref{eq:NSS4-1}-\eqref{eq:NSS4-2} into Eq.~\eqref{eq:NSS4} gives the three-term recurrence relation
\begin{equation}
P_{n+2}(\alpha,\xi)-\alpha P_{n+1}(\alpha,\xi) +  (n+1) \xi P_{n}(\alpha,\xi)=0 \, , \quad \, n=0,1,2,\ldots,
\label{eq:NSS5}
\end{equation}
along with the constraint
\begin{equation}
P_{1}(\alpha,\xi)-\alpha P_{0}(\alpha,\xi)=0.
\label{eq:NSS5-x}
\end{equation}
Equation (\ref{eq:NSS5-x}) becomes redundant if both $P_0$ and $P_1$ are equal to zero. Consistently, $P_0 =0$ is forbidden as initial condition since this produces $P_1=0$ in (\ref{eq:NSS5-x}), and leads to the trivial solution of  (\ref{eq:NSS5}). That is, the constraint (\ref{eq:NSS5-x}) induces a set $\{ P_n \}$ with $P_0 \neq 0$ by necessity. Therefore, if both equations (\ref{eq:NSS5}) and (\ref{eq:NSS5-x}) must be satisfied, the superposition (\ref{eq:NSS4-1}) is characterized by the probability amplitude $\mathcal{C}_0 \neq 0$.

Remark that the nonclassicality of $\vert \alpha, \xi \rangle$ may be limited by the classicalness of the vacuum state $\vert 0 \rangle$. The latter because $P_0 \neq 0$ means that $\vert 0 \rangle$ is always included in the profile of the superposition (\ref{eq:NSS4-1}), so the probability density $\vert \mathcal{C}_0 \vert^2 \neq 0$ could master the predictions for detecting $n$-photons in $\vert \alpha, \xi \rangle$. 

A different way to produce nonclassical states consists in dropping $\vert 0 \rangle$ from the superposition (\ref{eq:NSS4-1}), see for instance \cite{Aga91,Zav04,Zel16,Zel17a,Zel20}. As this means to take $P_0=0$, the complementary equation (\ref{eq:NSS5-x}) is not useful anymore and the eigenvalue equation (\ref{eq:NSS4}) is not necessarily satisfied. That is, preserving the recurrence relation (\ref{eq:NSS5}) to define the probability amplitudes (\ref{eq:NSS4-2}), but omitting to impose the constraint (\ref{eq:NSS5-x}) at the very beginning, we are in position to construct more general superpositions (\ref{eq:NSS4-1}). After that, the states $\vert \alpha, \xi \rangle$ so constructed would be asked to satisfy the appropriate initial conditions, the ones that may be ruled by (\ref{eq:NSS5-x}) in particular. We shall proceed in this form to investigate new possibilities of constructing nonclassical minimum uncertainty states.

Our approach is based on the calculus of finite differences \cite{Mil33}. We identify the three-term recurrence relation (\ref{eq:NSS5}) with a second-order difference equation and face it  by finding the general solution, the initial conditions of which are to be determined. Since this kind of difference equations admit two independent solutions \cite{Mil33}, we already know that one of such solutions will reproduce the well-known results associated to the system (\ref{eq:NSS5})-(\ref{eq:NSS5-x}). The other independent solution is therefore at our disposal to fix a different set of initial conditions such that the superposition (\ref{eq:NSS4-1}) is well defined. For details see Appendices~\ref{ApA} and \ref{ApB}. A second step is to determine whether these new states of light minimize a given uncertainty relationship.

We start by reviewing the values of $\alpha$ and $\xi$ that reduce the recurrence relation (\ref{eq:NSS5}) to the two-term case. Well-known minimum uncertainty states of light are recovered at the time that new squeezed states are introduced. Then we construct the general solution for the three-term case with arbitrary values of $\alpha$ and $\xi$.

\subsection{Special cases (two-term recurrence relations)}

Making either $\alpha =0$ or $\xi =0$, the recurrence relation (\ref{eq:NSS5}) is reduced to a two-term relationship. We analyze these cases separately.

$\bullet$ {\bf Glauber states.} For $\xi=0$ and $\alpha\in\mathbb{C}$ the eigenvalue problem (\ref{eq:NSS4}) is reduced to the Glauber's fully coherence condition $a \vert\alpha \rangle = \alpha \vert \alpha \rangle$. Indeed, the recurrence relation \eqref{eq:NSS5} is simplified to the first-order difference equation with constant coefficients:
\be
P_{n+2}(\alpha)-\alpha P_{n+1} (\alpha)=0, \quad n =0,1, \ldots
\label{arec}
\ee
Equation (\ref{arec}) may be reduced to $P_{n+1} = \alpha^n P_1$, and has the general solution $P_{n}(\alpha)= \kappa \alpha^{n}$, with $\kappa$ a constant fixed by the initial condition $P_0(\alpha) = \kappa$. Notice that the constraint (\ref{eq:NSS5-x}) is automatically satisfied since $P_1(\alpha) = \kappa \alpha$ holds for any value $\kappa$ of the initial condition. Without loss of generality, we take $\kappa=1$ in the present case. Therefore one obtains the Glauber states
\begin{equation}
\vert \alpha \rangle = e^{-\vert\alpha\vert^{2}/2}\sum_{n=0}^{\infty}\frac{\alpha^{n}}{\sqrt{n!}}\vert n \rangle, \quad  \alpha\in\mathbb{C}.
\label{Glauber}
\end{equation}
Although the coherent states $\vert \alpha \rangle$ are classical in the sense introduced by Glauber, unexpected properties were first noticed by Dodonov, Malkin and Man'ko \cite{Dod74}. They realized that the superpositions $\vert \alpha_{\pm} \rangle = c_{\pm} (\vert \alpha \rangle \pm \vert -\alpha \rangle)$, called {\em even} ($+$) and {\em odd} ($-$) coherent states, exhibit a nonclassical profile. Thus, whereas $\vert \alpha \rangle$ are classical states of light, the superpositions $\vert \alpha_{\pm} \rangle$ are not classical anymore~\cite{Dod74,Ger93}.

$\bullet$ {\bf Squeezed-vacuum states.} For $\alpha=0$ and $\xi \in \mathbb{C}$, we get the second-order difference equation with variable coefficients
\begin{equation}
P_{n+2}(\xi)+\xi(n+1)P_{n}(\xi)=0 \, , \quad n=0,1,2, \ldots
\label{eq:NSS4-3}
\end{equation}
The general solution of this equation includes two independent solutions. One of them defines a class of $P$-functions that satisfy the constraint (\ref{eq:NSS5-x}) as follows
\begin{equation}
P_{0}(\xi)=1, P_{1}(\xi)=0 \quad \Rightarrow \quad P_{2n}(\xi)=\frac{(2n)!}{2^n n!}(-\xi)^{n}, \quad P_{2n+1}(\xi)=0, 
\label{eq:NSS4-4}
\end{equation}
As the $P$-functions \eqref{eq:NSS4-4} are labeled by nonnegative integers $2n$, the state (\ref{eq:NSS4-1}) consists entirely of even-photon Fock state superpositions. We write
\begin{equation}
\vert \xi; + \rangle = (1-\vert\xi\vert^2)^{\frac{1}{4}}\sum_{n=0}^{\infty}\frac{\sqrt{(2n)!}}{2^n n!}(-\xi)^{n}\vert 2n \rangle, \quad \vert \xi \vert <1.
\label{eq:NSS5-1}
\end{equation}
The above expression defines the squeezed-vacuum state in non-unitary form, see for example \cite{Wun17}. Indeed, as $\vert n \rangle = (n!)^{-1/2} a^{\dagger n} \vert 0 \rangle$, we may use the disentangling formula
\be
S(\xi)= \exp \left\{ \frac{\operatorname{arctanh} \vert \xi \vert}{ 2 \vert \xi \vert} (\xi^* a^2 - \xi a^{\dagger 2}) \right\} = e^{-\frac{\xi}{2} a^{\dagger 2}} (1 - \vert \xi \vert^2)^{\frac14 (a a^{\dagger} + a^{\dagger} a)} e^{\frac{\xi^*}{2} a^2},
\label{disen1}
\ee
to write $\vert \xi; + \rangle = S(\xi) \vert 0 \rangle$, where the invariance of $\vert 0 \rangle$ under the action of $\exp( a^2)$ has been used. The proper selection of the parameter $\xi$ yields the single-mode squeezing operator (\ref{disen1}) in any of its well known representations.

$\bullet$ {\bf Odd-photon squeezed states.} The second independent solution of Eq.~(\ref{eq:NSS4-3}) satisfies initial conditions that are not ruled by the constraint (\ref{eq:NSS5-x}). The new family of $P$-functions is defined as follows
\begin{equation}
P_{0}(\xi)=0, P_{1}(\xi)=1 \quad \Rightarrow \quad  P_{2n+1}(\xi)=2^{n}n!(-\xi)^{n}, \quad P_{2n}(\xi)=0. 
\label{eq:NSS4-5}
\end{equation}
Then, the superpositions (\ref{eq:NSS4-1}) include odd-photon Fock states only. Explicitly,
\begin{equation}
\vert \xi ; - \rangle = \left( 1- \vert \xi \vert^2 \right)^{1/4} \left[ \frac{ \vert \xi \vert }{ \arcsin ( \vert \xi \vert ) } \right]^{1/2} \sum_{n=0}^{\infty} \frac{2^{n}n!}{\sqrt{(2n+1)!}}(-\xi)^{n}\vert 2n+1 \rangle, \quad \vert \xi \vert <1.
\label{eq:NSS5-2}
\end{equation}
Proceeding in a similar way to the previous case one arrives at the expression
\be
\vert \xi ; -\rangle = \frac{1}{\mathcal{N} (\xi; -)} \sum_{n=0}^{\infty} \frac{n!}{(2n+1)!} \left( - 2 \xi a^{\dagger 2} \right)^n \vert 1 \rangle,
\label{1phot}
\ee
where we have retrieved the symbolic form of the normalization factor for simplicity. The straightforward calculation shows that the above result admits further simplification
\be
\vert \xi; - \rangle = \frac{1}{\mathcal{N} (\xi; -)} \left[ \sum_{n=0}^{\infty} \frac{ \left( \frac12 \xi a^{\dagger 2} \right)^n}{n! (2 n + 1)} \right] \exp \left( - \frac{\xi}{2} a^{\dagger 2} \right) \vert 1 \rangle.
\label{1photb}
\ee
As the series in square brackets is a confluent hypergeometric function, we formally write
\be
\vert \xi;- \rangle = \mathcal{N}^{-1} (\xi; -) {}_1F_1 \left( \tfrac12, \tfrac32, \tfrac12 \xi a^{\dagger 2} \right) S(\xi)  \vert 1 \rangle,
\label{hyper}
\ee
where (\ref{disen1}) and the invariance of $\vert 1 \rangle$ under the action of $\exp( a^2)$ have been used. 

Following the terminology for the so called Hermite polynomial squeezed states\cite{Ber91,Dat96},  Laguerre polynomial states \cite{Fan95}, and E-exponential states \cite{Zel17a}, the odd-photon squeezed solutions (\ref{eq:NSS5-2})-(\ref{hyper}) may also be referred to as {\em squeezed confluent-hypergeometric} states (the term suggested by the relation ${}_1F_1(\tfrac12, \tfrac32, x^2) = \frac{\sqrt{\pi} }{2x}\operatorname{Erfi}(x^2)$ sounds a little bizarre). 

Quite interestingly, the states defined by (\ref{eq:NSS5-2})-(\ref{hyper}) are non-finite superpositions of the Pleba\'nski squeezed number states $\vert n, \xi \rangle_{\operatorname{Pleb}} = S(\xi) \vert n \rangle$ \cite{Ple54,Nie97}; the latter including the squeezed-vacuum $\vert \xi; + \rangle = S(\xi) \vert 0 \rangle$ as particular case. On the other hand, $\vert \xi;- \rangle$ is far from being an excitation of $\vert \xi ;+\rangle$ because $( a + \xi a^{\dagger})^{\dagger} \vert \xi ;+\rangle = (1 -\vert \xi \vert^2) S(\xi) \vert 1 \rangle$. 

To analyze the properties of the odd-photon squeezed states $\vert \xi; -\rangle$ we may use the Wigner function \cite{Wig32}, which enables the detection of squeezed states \cite{Wei18} and permits identifying nonclassicality in the quantum states \cite{Hil84}. Negative values of the the Wigner function betray the presence of non-Gaussian properties, which signifies {\em quantumness} \cite{Wig32,Hil84}. For our purposes we use the most general expression \cite{Aga91}
\begin{equation}
W(z)=e^{2\vert z\vert^{2}}\int_{\mathbb{C}}\frac{d^{2}\beta}{\pi}\langle -\beta \vert \rho\vert\beta\rangle e^{2(z\beta^{*}-z^{*}\beta)},
\label{eq:NCP2}
\end{equation}
with $\rho$ the density operator of the state under consideration, and the integration is being performed over the complex $\beta$-plane. The complex parameters $\beta$ and $z$ define respectively the Glauber coherent states $\vert\beta\rangle$ and $\vert z\rangle$. Our algorithm for depicting the Wigner function follows \cite{Zel18a,Zel20} with the pure state  $\rho_- = \vert \alpha,\xi; - \rangle\langle \alpha,\xi; -\vert$.

\begin{figure}[htb]
\centering
\subfloat[][$\xi=0.4$]{\includegraphics[width=0.3\textwidth]{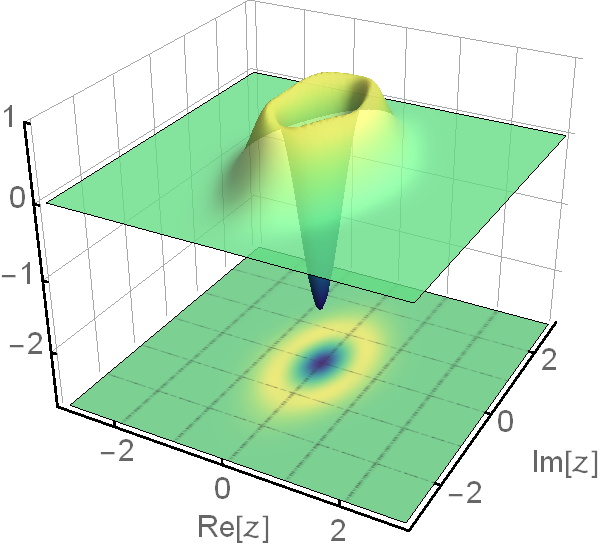}}
\hskip2ex
\subfloat[][$\xi=0.6$]{\includegraphics[width=0.3\textwidth]{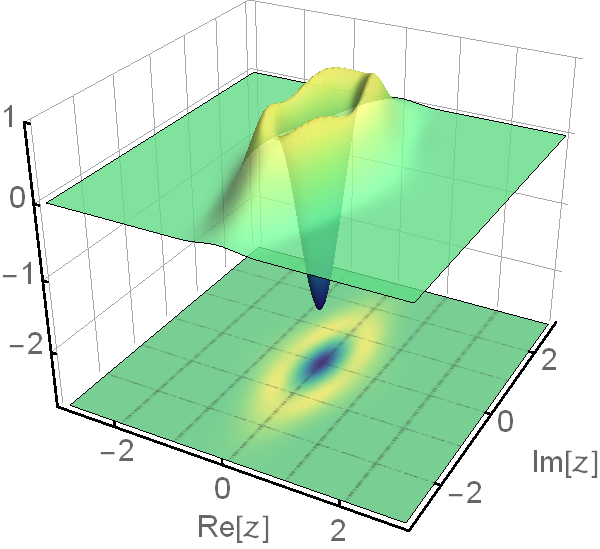}}
\hskip2ex
\subfloat[][$\xi=0.8$]{\includegraphics[width=0.3\textwidth]{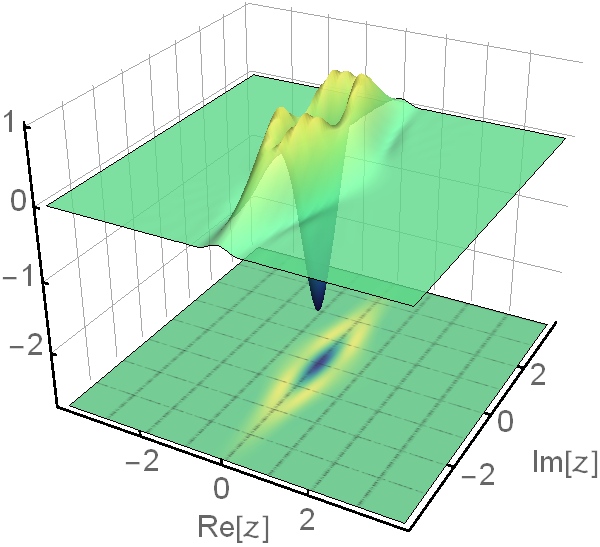}}
\caption{\footnotesize Wigner function defined by the odd-photon squeezed state $\vert \xi; -\rangle$ introduced in (\ref{eq:NSS5-2}) for the indicated values of the squeezing parameter $\xi$. The function is plotted on the complex $z$-plane characterizing the Glauber state $\vert z \rangle$ of Eq.~(\ref{eq:NCP2}). As $\xi \in \mathbb R$, the position variable is squeezed while the momentum variable is stretched, see Figure~\ref{Fsqueezing}.
}
\label{fig:F2}
\end{figure}

\begin{figure}[htb]
\centering
\subfloat[][$\xi=0$]{\includegraphics[width=0.24\textwidth]{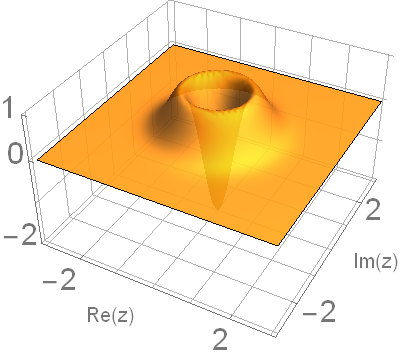}}
\hskip1ex
\subfloat[][$\xi=0.4 e^{i\pi/2}$]{\includegraphics[width=0.24\textwidth]{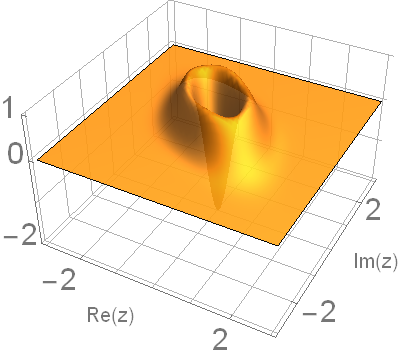}}
\hskip1ex
\subfloat[][$\xi=0.4 e^{i\pi}$]{\includegraphics[width=0.24\textwidth]{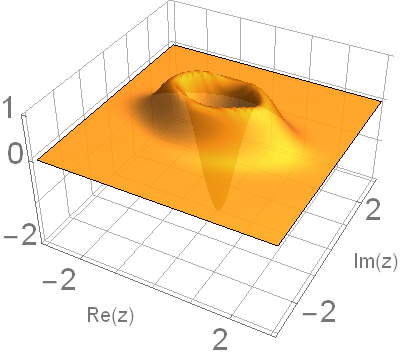}}
\hskip1ex
\subfloat[][$\xi=0.4 e^{3i\pi/2}$]{\includegraphics[width=0.24\textwidth]{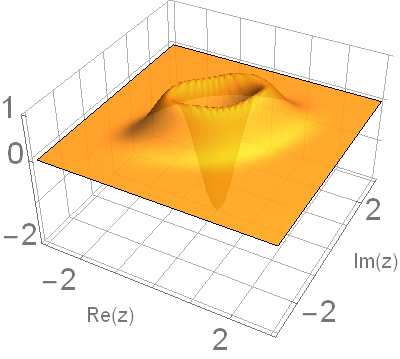}}
\caption{\footnotesize Wigner function defined by the odd-photon squeezed state $\vert \xi; -\rangle$ introduced in (\ref{eq:NSS5-2}) for the indicated values of the complex parameter $\xi = \vert \xi \vert \exp (i\theta_{\xi})$. Compare with Figure~\ref{fig:F2}.
}
\label{FigW2}
\end{figure}

We illustrate the Wigner function of state $\vert \xi; -\rangle$ in Figure~\ref{fig:F2} for three different real values of the parameter $\xi$. Before starting our analysis, we would like to emphasize the result 
\be
\lim_{\vert \xi \vert \rightarrow 0} \vert \xi; - \rangle = \vert 1 \rangle,
\ee
which is easily verified by inspecting Eq.~(\ref{eq:NSS5-2}). That is, cancelling the  squeezing parameter $\xi$ in the odd-photon squeezed state $\vert \xi; -\rangle$ we arrive at the one-photon Fock state $\vert 1 \rangle$. Recalling that $\vert \xi; -\rangle$ is defined on the open disk of unit radius $\vert \xi \vert <1$ in the complex $\xi$-plane, we see that the deformation suffered by the Wigner function signifies a transition of the quantum profile of $\vert \xi; - \rangle$ that never arrives at the classical limit. The latter is in clear contraposition to the one-photon added coherent state $\vert \alpha, 1 \rangle_{\operatorname{add}}$ for which the quantum-to-classical transition can be measured and characterized by quantum tomography \cite{Zav04}. Indeed, at $\xi =0$ the Wigner function for $\vert \xi; -\rangle$ is just the same as that for $\vert 1 \rangle$, so the quantumness of $\vert 1 \rangle$ is also associated to $\vert \xi =0; -\rangle$, see Figure~\ref{FigW2}. Nevertheless, as soon as $\xi$ is different from zero, we find squeezing in the $x$ quadrature. The higher the value of $\xi \in \mathbb R$, the stronger the squeezing of $x$. In other words, the state $\vert \xi; - \rangle$ is never classical for $\xi \in \mathbb R$, but it may be either squeezed $(\xi \neq 0$) or not squeezed $(\xi = 0$). In the more general case $\xi \in \mathbb C$, the Wigner functions configured with real $\xi$ are affected just by a rotation.

Up to our knowledge, the states $\vert \xi ;-\rangle$ have not been reported anywhere before the present work. Whereas the vectors $\vert \xi ;+ \rangle$ were brought to light due to practical necessities in the interferometry of gravitational-waves, the odd-photon squeezed states $\vert \xi ;- \rangle$ seem to be even missing in theoretical approaches dealing with minimum uncertainty states. We think that the lack of results on this subject is because most of the   approaches omit the deep revision of the corresponding recurrence relations. 

The above note does not apply to the derivation of the Glauber states (\ref{Glauber}) since the recurrence relationship is a first-order difference equation, so it admits only one independent solution. The even and odd coherent states introduced by Dodonov, Malkin and Man'ko \cite{Dod74} are in this respect different expressions of the same solution. 


\section{Solutions to the three-term recurrence relation}
\label{section3}

In this section we provide solutions to the recurrence equation (\ref{eq:NSS5}) by following the comparison method introduced in Appendix~\ref{ApA}. We proceed by finding a first solution of Eq.~(\ref{eq:NSS5}), which will reproduce results already reported in the literature after considering the constraint (\ref{eq:NSS5-x}). Then, using the formulation included in Appendix~\ref{ApB}, we find a second independent solution for (\ref{eq:NSS5}) and show that this does not fulfill the constraint (\ref{eq:NSS5-x}). Nevertheless, such a solution is well-defined and gives rise to new forms of squeezed states.

\subsection{Conventional two-parameter squeezed states}

A first solution of the three-term recurrence relation \eqref{eq:NSS5} is easily achieved and leads to the family (see details in Appendix~\ref{ApA}):
\be
P_n(\alpha, \xi; +) = \left(\tfrac{\xi}{2} \right)^{\frac{n}{2} }H_n \left( \tfrac{\alpha}{\sqrt{2 \xi} } \right)= \xi^{\frac{n}{2}} H\!e_n \left( \tfrac{\alpha}{\sqrt{\xi} } \right)
,
\label{2herm}
\ee 
with $H_n(x) =(-1)^n e^{x^2} \frac{d^n}{dx^n} e^{-x^2}$ and $H\!e_n (x) = 2^{-\frac{n}{2}} H_n \left(\tfrac{x}{\sqrt 2} \right)$ the Hermite and {\em scaled} Hermite polynomials \cite{Olv10}, respectively. 

Strikingly, $P_n(\alpha, \xi; +)$ behaves like the scaled Hermite polynomial $H\!e_{n} (\alpha)$ in the parameter $\alpha$, see Table~\ref{tableH}. In particular, from (\ref{2herm}) we find $P_n(\alpha, \xi = 1; +)= H\!e_{n} (\alpha)$, so that $P_n(\alpha, \xi; +)$ is a polynomial of order $n$ in $\alpha$ for $\xi=1$. On the other hand, inspecting Table~\ref{tableH} we realize that $P_n(\alpha, \xi; +)$ is a polynomial of degree $\lfloor n/2 \rfloor$ in $\xi$, with $\lfloor z \rfloor$ the floor function of $z$. It is then remarkable that $P_n(\alpha, \xi; +)$ is twice degenerate in the $n$th order of $\xi$.

\begin{table}
\centering
\begin{tabular}{cccc}
\hline
n & $H_n(x)$ & $H_{en}(x)$ & $P_{n}(\alpha,\xi; +)$  \\
\hline
0 & 1 &1 & 1\\[.5ex]
1 & $2x$ & $x$ & $\alpha$  \\[.5ex]
2 & $4x^2 -2$ & $x^2-1$ &$\alpha^2-\xi$  \\[.5ex]
3 & $8x^3 -12 x$ & $x^3 -3x$ & $\alpha^3-3\xi\alpha$ \\[.5ex]
4 & $16x^4 -48 x^2 +12$ & $x^4 -6x^2 +3$ & $\alpha^4-6\xi\alpha^2+3\xi^2$ \\
\hline
\end{tabular}
\caption{\footnotesize First five elements of the Hermite polynomials $H_n(x)$ and $H_{en}(x)$. They are compared with the two-parametric Hermite polynomials $P_{n} (\alpha, \xi; +)$ introduced in Eq.~(\ref{2herm}). 
}
\label{tableH}
\end{table}

Hereafter, the $P$-functions (\ref{2herm}) will be referred to as {\em two-parametric} Hermite polynomials. Notice that this family satisfies $P_0 =1$ and $P_1=\alpha$, so the constraint (\ref{eq:NSS5-x}) is fulfilled. Therefore,
\begin{equation}
\vert\alpha,\xi ;+ \rangle = (1-\vert\xi\vert^{2})^{\frac{1}{4}} e^{-\frac{ \vert \alpha \vert^{2}}{2 (1-\vert \xi \vert^{2})}} e^{\frac{ ( \xi \alpha^{*2}+ \xi^{*} \alpha^{2} ) }{4 (1-\vert \xi \vert^{2} )}} 
\sum_{n=0}^{\infty}\frac{ \xi^{\frac{n}{2}} }{ \sqrt{n!}} H\!e_{n} \left( \tfrac{\alpha}{\sqrt{\xi}}\right) 
\vert n \rangle, \quad \vert \xi \vert < 1.
\label{eq:NSS6}
\end{equation}
The normalization factor (\ref{factor}) for these vectors has been achieved through the Mehler's formula \cite{Erd53} (see Sec.~10.13, p.~194). 

The two-parametric minimum uncertainty states defined in \eqref{eq:NSS6} reproduce the usual expression of the conventional squeezed states \cite{GrupoS} for the appropriate profile of the $\xi$-parameter. In particular, the squeezed-vacuum $\vert \xi; + \rangle$ is recovered at the limit $\vert \alpha \vert \rightarrow 0$. 

\subsection{Expanding the set of squeezed states}

Provided $P_n(\alpha, \xi; +) $, the second independent solution of the recurrence relation (\ref{eq:NSS5}) yields the family
\begin{equation}
P_{n+1} (\alpha,\xi ;-) = \left(\frac{\xi}{2}\right)^{\frac{n}{2}} \hspace{-2mm} H_{n+1}(y)\sum_{k=0}^{n} \frac{2^{k}k!}{H_{k}(y)H_{k+1}(y)}, \quad  y = \frac{\alpha}{\sqrt{2 \xi}}, \quad  n = 0, 1, \ldots,
\label{eq:CSSS1}
\end{equation}
see details in Appendices~\ref{ApA} and \ref{ApB}. The straightforward calculation shows that (\ref{eq:CSSS1}) can be rewritten in the form
\be
P_{n+1} (\alpha, \xi; -) =\left( \tfrac{\xi}{2} \right)^{\frac{n}{2}} H_n^{\nu =1} \left( \tfrac{\alpha}{\sqrt{2 \xi}} \right) = \xi^{\frac{n}{2}}H\!e_n^{\nu=1} \left( \tfrac{\alpha}{\sqrt{\xi}} \right),
\label{Wuns1}
\ee
where $H_n^{\nu}(x)$ stands for the {\em associated Hermite polynomials} \cite{Ask84,Mou05} as they are revisited in \cite{Wun03,Wun19}, compare with Eq.~(\ref{2herm}).

The $P$-functions (\ref{eq:CSSS1})-(\ref{Wuns1}) do not satisfy the constraint (\ref{eq:NSS5-x}) since their derivation requires the initial value $P_{0} (\alpha,\xi; - )=0$. Indeed, they satisfy the constraint 
\be
P_2(\alpha, \xi; -) - \alpha P_1(\alpha, \xi; -)  =0.
\label{distorted}
\ee
Comparing this result with (\ref{eq:NSS5-x}) could lead to the wrong conclusion that (\ref{distorted}) is a shifted version of the constraint obeyed by the squeezed states $\vert \alpha, \xi; + \rangle$. However, the main difference is that $P_0(\alpha, \xi; -)=0$ whereas $P_0(\alpha, \xi; +)=1$, see Table~\ref{tab:T1}. The remaining $P$-functions are different from zero in both cases, with $P_{n+1}(\alpha, \xi; -)$ and $P_n(\alpha, \xi; +)$ sharing the same order in $\alpha$ and $\xi$. Thus, $P_n(\alpha, \xi; -)$ is twice degenerate in the zero order of $\alpha$ and twice degenerate in the $n$th order of $\xi$.

\begin{table}
\centering
\begin{tabular}{ccc}
\hline
n & $P_{n}(\alpha,\xi; +)$ & $P_{n} (\alpha, \xi; -)$ \\
\hline
0 & 1 &  0 \\[.5ex]
1 & $\alpha$ &  1 \\[.5ex]
2 & $\alpha^2-\xi$ &  $\alpha$ \\[.5ex]
3 & $\alpha^3-3\xi\alpha$ &  $\alpha^2-2\xi$ \\[.5ex]
4 & $\alpha^4-6\xi\alpha^2+3\xi^2$ &  $\alpha^3-5\xi\alpha$ \\
\hline
\end{tabular}
\caption{\footnotesize First five elements of the classical polynomials $P_{n} (\alpha, \xi; +) =  (\tfrac{\xi}{2})^{\frac{n}{2} }H_n ( \frac{\alpha}{\sqrt{2 \xi} })$ and the associated ones $P_{n+1} (\alpha, \xi; -) =  (\tfrac{\xi}{2})^{\frac{n}{2} }H_n^{\nu=1} ( \frac{\alpha}{\sqrt{2 \xi} })$, with $P_0(\alpha, \xi; -)=0$. Note that $P_{n} (\alpha, \xi; -)$ is twice degenerate in the zero-order case.
}
\label{tab:T1}
\end{table}

Next some basic properties of the associated Hermite polynomials $P_{n} (\alpha, \xi; -) $ are reviewed in comparison with the classical ones $P_{n} (\alpha, \xi; +) $.
 
\subsubsection{Properties of the associated Hermite polynomials}

Both $P$-functions, $P_n (\alpha, \xi; +) $ and  $P_{n+1} (\alpha,\xi; - )$, are polynomials of degree $n$ in $\alpha$ and degree $\lfloor n/2 \rfloor$ in $\xi$. Both sets are free of singularities and such that their leading coefficient is equal to 1 for any $n$. That is, they are monic polynomials of degree $n$ in $\alpha$. 

\begin{figure}[htb]
\centering
\subfloat[][$P_{n}(\alpha, \xi = 0.6; +)$]{\includegraphics[width=0.4\textwidth]{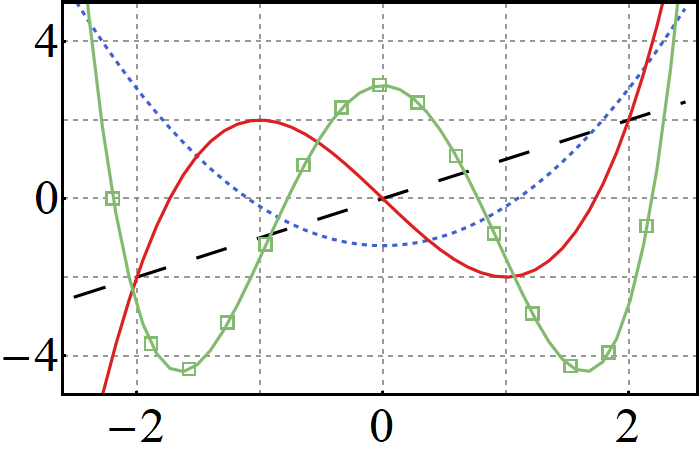}
}
\hskip5ex
\subfloat[][$P_{n+1}(\alpha, \xi = 0.6; -)$]{\includegraphics[width=0.4\textwidth]{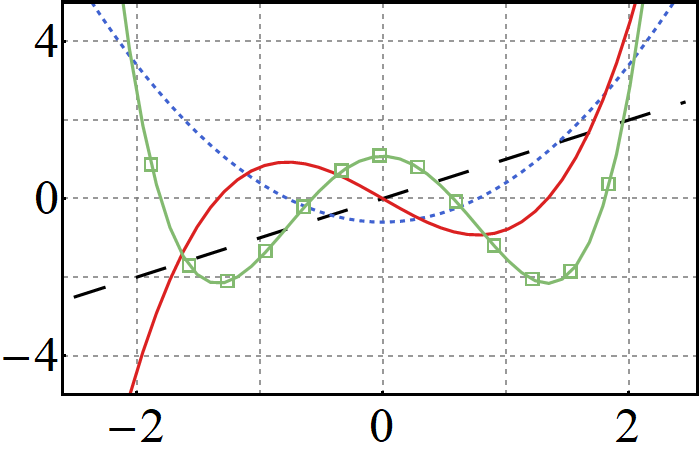}
}
\caption{\footnotesize Polynomials $P_{n}(\alpha, \xi; +)$ and $P_{n+1} (\alpha, \xi; -)$ as a function of $\alpha\in\mathbb{R}$ for the real parameter $\xi=0.6$, and $n=1$ (long-dashed-black), $n=2$ (dotted-blue), $n=3$ (solid-red) and $n=4$ (squared-green). See the  analytical form of these polynomials on Table~\ref{tab:T1}.
}
\label{fig:Pol}
\end{figure}

The pairing of orders between $P_n (\alpha,\xi; + )$ and $P_{n+1} (\alpha,\xi ;- )$ means a shift in the distribution of zeros for these families of polynomials, see Figure~\ref{fig:Pol}. In both cases the interlacing of zeros obeys an oscillation rule that mimics the one satified by real-valued polynomials. In the Hermitian case this means orthogonality and vice versa. Nevertheless, the latter does not apply to the non-Hermitian case since complex-valued polynomials are not orthogonal in the conventional form.The detailed analysis of such a property is out of the scope of the present work. In any case, the Favard's theorem \cite{Chi78,Fav35} may be useful on that matter.

To continue our analysis we first rewrite the $P$-functions (\ref{eq:CSSS1})-(\ref{Wuns1}) as a power series in $\alpha$, then we decouple the result into even and odd contributions. After some simplifications (the material included in Appendix \ref{ApA} is very useful on the matter), we arrive at the expressions
\be
\begin{array}{c}
P_{2n+2} (\alpha,\xi; - ) =  c_n(\alpha,\xi)
\displaystyle\sum_{m=0}^{n} \left(\tfrac{\alpha^{2}}{2\xi}\right)^{m} 
\frac{(-n)_{m}}{(\tfrac32)_{m}(m+1)!}
\,  {}_{3}F_{2}\left( \left. 
\begin{aligned} 
1, \tfrac12,m-n \\[1ex] 
m+2, m+ \tfrac32  
\end{aligned} \, 
\right\vert 1 \right) ,
\end{array}
\label{eq:CSSS1-2}
\ee
with $c_n(\alpha,\xi)= \alpha (n+1) (\tfrac32)_{n}(-2\xi)^{n}$, and
\begin{equation}
\begin{alignedat}{3}
& P_{2n+1} (\alpha,\xi; - ) = ( \tfrac32)_{n} (-2\xi)^{n} \sum_{m=0}^{n}\left( \tfrac{\alpha^{2}}{2\xi}\right)^{m} \frac{(-n)_{m}}{(\tfrac32)_{m}m!} 
\, {}_{3}F_{2}\left( \left. 
\begin{aligned} 
1, \tfrac12 ,m-n \\[1ex] 
m+1,m+ \tfrac32 
\end{aligned} \, \right\vert 1 \right).
\end{alignedat}
\label{eq:CSSS1-3}
\end{equation}
Here ${}_q F_p (\cdots \vert z)$ is the generalized hypergeometric function, $(a)_{n}=\Gamma(a+n)/\Gamma(a)$ the Pochhammer symbol, and  $\Gamma(z)$ the gamma function \cite{Olv10}. 

With the above results it is easy to verify the following conclusions:

i) At the limit $\alpha \rightarrow 0$ we have $\left. P_{2n+2} (\alpha,\xi; - )\right\vert_{\alpha=0} =0$ for any $n$. In turn, the odd contribution $\left. P_{2n+1} (\alpha,\xi; - )\right\vert_{\alpha=0} = 2^n n!  (-\xi)^n$ coincides with the $P$-functions defined for the odd-photon squeezed states $\vert \xi; - \rangle$ in Eq.~(\ref{eq:NSS4-5}).

ii) At the limit $\xi\rightarrow 0$ both polynomials $P_{2n+1}(\alpha, \xi; -)$ and $P_{2n} (\alpha, \xi; -)$ are different from zero. Besides, their dependence on $\alpha$ is free of singularities and the initial condition $P_{0} (\alpha, \xi; -)=0$ is preserved. 

\subsubsection{Associated squeezed states}
\label{asocia}

The introduction of \eqref{eq:CSSS1} into \eqref{eq:NSS4-1} yields
\begin{equation}
\vert \alpha,\xi; - \rangle = \frac{1}{\mathcal{N}(\alpha,\xi; -)} \sum_{n=0}^{\infty} \sum_{k=0}^n \left(\frac{\xi}{2}\right)^{n/2} \frac{H_{n+1}(y)}{\sqrt{n!}} \frac{2^{k}k!}{H_{k}(y)H_{k+1}(y)} \vert n+1 \rangle, \quad \vert \xi \vert <1.
\label{associated}
\end{equation}
Equivalently, using (\ref{Wuns1}) one obtains
\be
\vert \alpha,\xi; - \rangle = \frac{1}{\mathcal{N}(\alpha,\xi; -)}  \sum_{n=0}^{\infty} \frac{\xi^{\frac{n}{2}}}{\sqrt{(n+1)!}} H\!e_n^{\nu=1} \left( \tfrac{\alpha}{\sqrt{ \xi}} \right) \vert n+1 \rangle, \quad \vert \xi \vert <1,
\label{Wuns2}
\ee
compare with (\ref{eq:NSS6}).

Any of the expressions (\ref{associated}) or (\ref{Wuns2}) defines the {\em associated squeezed states}, a class of minimum uncertainty states that is complementary to the well established family of squeezed states $\vert \alpha,\xi; + \rangle$. As far as we know, the new set of squeezed states $\vert \alpha,\xi; - \rangle$ has not been reported anywhere before the present work. 

The statistical and nonclassical properties of the associated squeezed states(\ref{associated})-(\ref{Wuns2}) are discussed in the next sections. Before that, we specialize on some concrete cases that are immediately obtained from the expression (\ref{associated}).

$\bullet$ {\bf Recovering the odd-photon squeezed states.} From item (i) of the previous section we know that the even functions (\ref{eq:CSSS1-2}) do not contribute to the profile of $\vert \alpha, \xi; - \rangle$ at $\alpha =0$. Thus, none of the even-number Fock vectors $\vert 2 n \rangle$ is included in the superposition $\vert \alpha=0, \xi; - \rangle$, as this would be expected. On the other hand, as the polynomials $P_{2n+1} (\alpha,\xi; - )$ lead to the coefficients of the odd-photon squeezed state $\vert \xi; - \rangle$ at $\alpha =0$, we realize that only the odd-number Fock vectors $\vert 2n+1 \rangle$ contribute to the identity $\vert \alpha=0, \xi; - \rangle = \vert \xi; - \rangle$, as this was already devised. 

As it can be seen, the odd-photon squeezed states $\vert \xi; - \rangle$ introduced in Eq.~(\ref{eq:NSS4-5}) are associated squeezed states $\vert \alpha, \xi; - \rangle$ with $\alpha =0$.

 $\bullet$ {\bf Distorted coherent states.} From item (ii) of the previous section we know that both functions (\ref{eq:CSSS1-2})  and (\ref{eq:CSSS1-3}) contribute to the profile of $\vert \alpha, \xi; - \rangle$ at $\xi =0$. Besides, the condition $P_{0} (\alpha, \xi; -)=0$ means that the vacuum $\vert 0 \rangle$ is excluded from the superposition (\ref{associated}). Let us pay attention to the resulting vector
\begin{equation}
\left. \vert \alpha, \xi; - \rangle \right\vert_{\xi =0}\equiv \vert \alpha; - \rangle = \frac{\vert\alpha\vert}{\sqrt{e^{\vert\alpha\vert^{2}}-1}}\sum_{n=0}^{\infty}\frac{\alpha^{n}}{\sqrt{(n+1)!}}\vert n+1 \rangle,
\label{eq:CSS3}
\end{equation}
which is a {\em distorted coherent state} \cite{Fer95,Ros96} and exhibits nonclassical behavior. To clarify the point let us pay attention to the relationship
\be
a \vert \alpha; - \rangle=
\frac{\vert \alpha \vert}{\sqrt{1 - e^{-\vert \alpha \vert^2}}} \vert \alpha \rangle.
\label{1distor}
\ee
Thus, up to a constant factor, the action of the boson-annihilation operator $a$ on the distorted coherent state $\vert \alpha; - \rangle$ produces the coherent state $\vert \alpha \rangle$, which is classical. This result resembles what happens when the operator $a$ acts on the 1-photon number state $\vert 1 \rangle$, resulting in the vacuum state $\vert 0 \rangle$. One may say that $\vert \alpha; - \rangle$ is nonclassical as compared with the Glauber state $\vert \alpha \rangle$.

We can go a step further by considering the expression for the one-photon added coherent state  \cite{Aga91}:
\begin{equation}
\vert\alpha,1\rangle_{\operatorname{add}} = 
\frac{e^{-\vert\alpha\vert^2/2}}{\sqrt{1+\vert\alpha\vert^2}}\sum_{n=0}^{\infty} \left( \frac{n+1}{n!} \right)^{1/2} \!\! \alpha^{n} \vert n+1\rangle.
\label{eq:CSS4}
\end{equation}
It is well known that $\vert\alpha,1 \rangle_{\operatorname{add}}$ results from the action of the boson-creation operator $a^{\dagger}$ on the Glauber state $\vert \alpha \rangle$. To compare (\ref{eq:CSS3}) with (\ref{eq:CSS4}) let us apply $a^{\dagger}$ on Eq.~(\ref{1distor}), we get
\be
\hat n \vert \alpha;- \rangle=
\frac{ \vert \alpha \vert \sqrt{1 + \vert \alpha \vert^2 }}{ \sqrt{1 - e^{-\vert \alpha \vert^2}}}
\, \vert\alpha,1\rangle_{\operatorname{add}} 
, \qquad \hat n = a^{\dagger} a.
\label{1photon}
\ee
Thus, the one-photon added coherent state $\vert\alpha,1\rangle_{\operatorname{add}}$ is the result of applying the number operator $\hat n$ on the distorted coherent state $\vert \alpha; - \rangle$. 

\begin{figure}[htb]
\centering
\subfloat[][$\vert \alpha, 1 \rangle_{\operatorname{add}}$\\$\alpha=e^{3i\pi/4}$]{\includegraphics[width=0.24\textwidth]{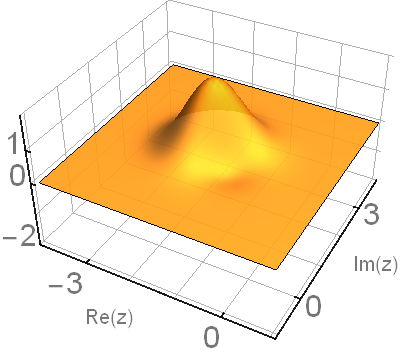}}
\hskip1ex
\subfloat[][$\vert \alpha; - \rangle$\\$\alpha=e^{3i\pi/4}$]{\includegraphics[width=0.24\textwidth]{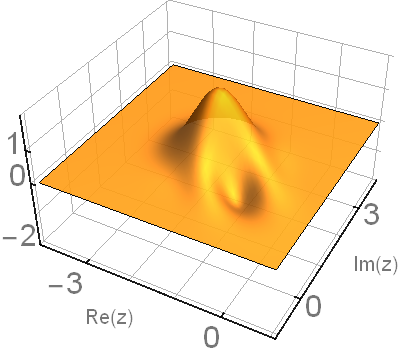}}
\hskip1ex
\subfloat[][$\vert \alpha, 1 \rangle_{\operatorname{add}}$\\$\alpha=2 e^{3i\pi/4}$]{\includegraphics[width=0.24\textwidth]{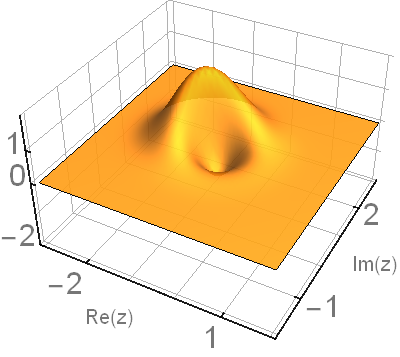}}
\hskip1ex
\subfloat[][$\vert \alpha; - \rangle$\\$\alpha=2 e^{3i\pi/4}$]{\includegraphics[width=0.24\textwidth]{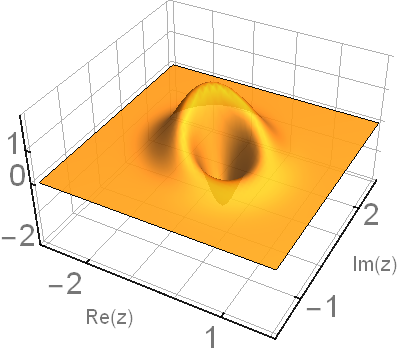}}
\caption{\footnotesize Wigner functions associated with the distorted coherent state $\vert \alpha; - \rangle$ and the one-photon added coherent state $\vert \alpha, 1 \rangle_{\operatorname{add}}$ for the indicated values of the complex parameter $\alpha = \vert \alpha \vert \exp( i\theta_{\alpha})$.
}
\label{FigW3}
\end{figure}

Figure~\ref{FigW3} shows the Wigner function of the one-photon added coherent state $\vert \alpha, 1 \rangle_{\operatorname{add}}$ and the distorted coherent state $\vert \alpha; - \rangle$ for two different values of $\alpha$. Although we identify zones with negative values of $W(z)$ for both states, the deformations of $W(z)$ for $\vert\alpha,-\rangle$ are much stronger than those generated by $\vert\alpha,1\rangle_{\operatorname{add}}$. Then, we say that $\vert\alpha,-\rangle$ is more nonclassical than $\vert \alpha, 1 \rangle_{\operatorname{add}}$. Besides, albeit both vectors represent nonclassical states of light, the relationship (\ref{1photon}) shows that the distorted coherent state $\vert \alpha; - \rangle$ can be considered the generating function of the one-photon added coherent state $\vert \alpha, 1 \rangle_{\operatorname{add}}$.

An additional result is easily derived from the formula
\be
\left[1 + \alpha \frac{d}{d\alpha} \right]  \vert \alpha;- \rangle =
\frac{ \vert \alpha \vert \sqrt{1 + \vert \alpha \vert^2 }}{ \sqrt{1 - e^{-\vert \alpha \vert^2}}}
\, \vert\alpha,1\rangle_{\operatorname{add}},
\label{nice}
\ee
which shows that $\vert \alpha;- \rangle$ provides a form to determine the Segal-Bargmann representation for the boson ladder operators in terms of the parameter $\alpha$ (see for instance \cite{Ros96} and references quoted therein).

\subsection{General structure of the space of solutions}
\label{structure}

To get a measure of the distinguishability of the pair $\vert\alpha,\xi; \pm \rangle$ we may use the trace norm, defined for any trace class operator $A$ as $\vert\vert A \vert\vert = \operatorname{tr} \vert A \vert$, with $\vert A \vert = \sqrt{A^{\dagger} A}$ the modulus of $A$ \cite{Nie00}. Therefore, $d(\rho_+, \rho_-) =\frac12 \vert \vert \rho_+ - \rho_- \vert \vert$ satisfies $0 \leq d(\rho_+, \rho_-) \leq 1$, with $d(\rho_+, \rho_-) =1$ if the states are distinguishable (orthogonal) and $d(\rho_+, \rho_-) =0$ for $\rho_+ =\rho_-$. 

\begin{figure}[htb]
\centering
\subfloat[][$d(\rho_{+},\rho_{-})$]{\includegraphics[width=0.3\textwidth]{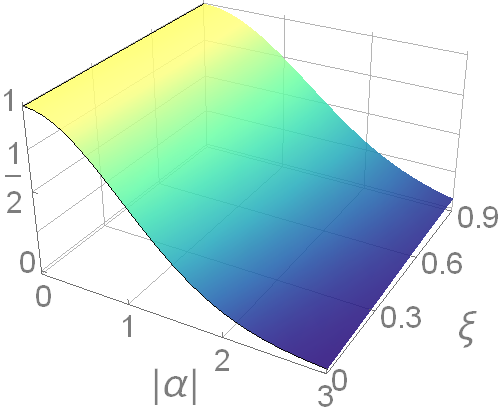}}
\hskip2ex
\subfloat[][$\langle N_+ \rangle$]{\includegraphics[width=0.3\textwidth]{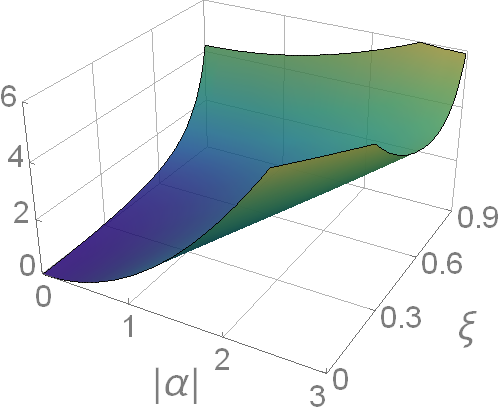}}
\hskip2ex
\subfloat[][$\langle N_- \rangle$]{\includegraphics[width=0.3\textwidth]{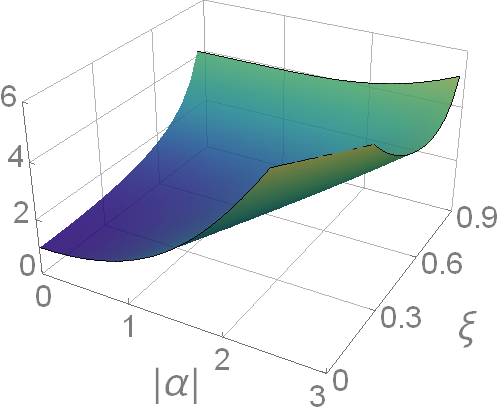}}

\caption{\footnotesize
{\bf (a)} Trace distance $d(\rho_+, \rho_-)$ between the squeezed state $\rho_+ = \vert\alpha,\xi; + \rangle \langle \alpha, \xi; + \vert$ and the associated squeezed state $\rho_- = \vert\alpha,\xi; - \rangle \langle \alpha, \xi; - \vert$ as a function of $\vert \alpha \vert$, with $\xi \in (0,1)$. These states are as distinguishable (orthogonal) as $\vert \alpha \vert \rightarrow 0$, for which $d(\rho_+, \rho_-) \rightarrow 1$. The values of $\vert \alpha \vert$ and $\xi$ producing $d(\rho_+, \rho_-) = 0$ are such that the vectors $\vert\alpha,\xi; \pm \rangle$ represent exactly the same squeezed state. Average number of photons $\langle N(\alpha, \xi; +) \rangle$ {\bf (b)} and $\langle  N(\alpha, \xi; -)  \rangle$ {\bf (c)} for the related states as a function of $\vert\alpha\vert$ and $\xi$.
}
\label{inner}
\end{figure}

Figure~\ref{inner}(a) shows the trace distance between the squeezed states $\rho_{\pm} = \vert\alpha,\xi; \pm \rangle \langle \alpha, \xi; \pm \vert$ for real values of $\xi$, and as a function of $\vert \alpha \vert$. Clearly, $d(\rho_+, \rho_-) =1$ at $\alpha=0$, meaning that the squeezed-vacuum $\vert \xi; + \rangle$ and the odd-photon squeezed state $\vert \xi; - \rangle$ are orthogonal, as this was anticipated in the previous sections. On the other hand, for $\vert \alpha \vert \neq 0$ the squeezed state $\vert\alpha,\xi; + \rangle$ and the associated squeezed state $\vert\alpha,\xi; - \rangle$ are less distinguishable as $\vert \alpha \vert \rightarrow \infty$ since $d(\rho_+, \rho_-)  \rightarrow 0$. At the very limit, $d(\rho_+, \rho_-) =0$ means that both vectors $\vert\alpha,\xi; \pm \rangle$ represent exactly the same squeezed state. The trace distance is therefore a measure that permits to delimitate the convergence radius at which these vectors may be assumed as undistinguishable. 

To associate the parameters $\alpha$ and $\xi$ with measurable quantities of the system let us calculate the average number of photons $\langle N (\alpha, \xi; \pm ) \rangle$ for the squeezed states $\vert \alpha, \xi; \pm \rangle$. We may proceed either by applying a Bogoliubov transformation on the ladder operators \cite{Nie97} or by calculating $\langle N (\alpha, \xi; \pm ) \rangle =\operatorname{Tr}(a^{\dagger}a\rho_{\pm})$ directly. In the latter case, for $\langle N (\alpha, \xi; +) \rangle$ we obtain a very simple expression
\begin{equation}
\langle N (\alpha, \xi; +) \rangle =\frac{\vert\xi\vert^{2}(1-\vert\xi\vert^{2})+\vert\alpha\vert^{2}(1+\vert\xi\vert^{2})-\alpha^{2}\xi^{*}-\alpha^{*2}\xi}{(1-\vert\xi\vert^{2})^{2}}.
\label{nplus}
\end{equation}
To obtain the above result, Eq.~5.12.2.1 of \cite{Pru86} is useful. Figure~\ref{inner}(b) shows the behavior of the average (\ref{nplus}).

In turn, it is not feasible to get a closed expression for $\langle N (\alpha, \xi; -) \rangle$ in general. Nevertheless, two particular cases are of remarkable interest. Namely,
\begin{equation}
\left. \langle N (\alpha, \xi; -) \rangle \right\vert_{\alpha=0} =\frac{\vert\xi\vert}{\arcsin(\vert\xi\vert)\sqrt{1-\vert\xi\vert^{2}}}+\frac{\vert\xi\vert^{2}}{1-\vert\xi\vert^{2}}, \quad \vert\xi\vert<1,
\label{nminusa}
\end{equation}
and
\begin{equation}
\left. \langle N (\alpha, \xi; -) \rangle \right\vert_{\xi=0} =\frac{\vert\alpha\vert^{2}}{1-e^{-\vert\alpha\vert^{2}}}.
\label{nminusb}
\end{equation}
Figure~\ref{inner}(c) shows the numerical calculation of $\langle N (\alpha, \xi; -) \rangle$, the boundaries of which are defined by $\alpha=0$ and $\xi=0$, see Eqs.~(\ref{nminusa}) and (\ref{nminusa}), respectively. Comparing the Figures~\ref{inner}(b) and \ref{inner}(c) one realizes that a given point $(\vert \alpha \vert, \xi)$, with $\xi \in (0,1)$, yields a different distribution of photons in each case. This means that the trace distance shown in Figure~\ref{inner}(a) links states with different averages in the number of photons for every  point $(\vert \alpha \vert, \xi)$.

\subsubsection{Probability distribution functions}

\begin{figure}[tbh]
\centering
\subfloat[][$\mathcal{P}_n (\alpha, \xi; +)$]{\includegraphics[width=0.4\textwidth]{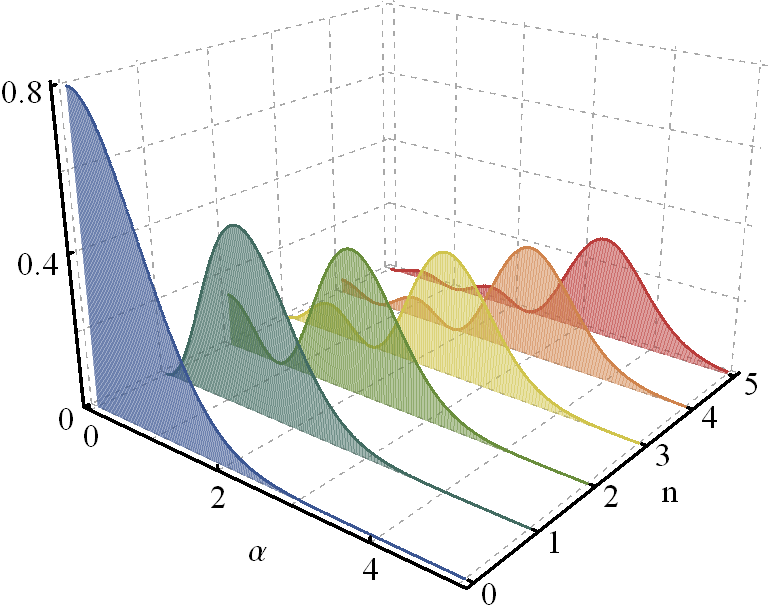}}
\hskip5ex
\subfloat[][$\mathcal{P}_n (\alpha, \xi; -)$]{\includegraphics[width=0.4\textwidth]{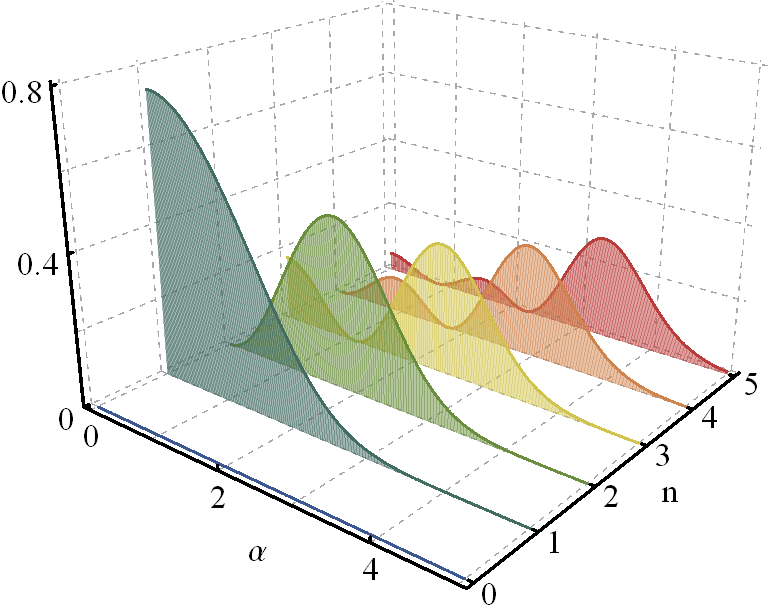}}
\caption{\footnotesize Probability distribution functions of the squeezed states $\vert \alpha, \xi; \pm \rangle$ for $\xi =0.6$. The probability of finding $n$-photons decreases as $\vert \alpha \vert \rightarrow \infty$. The number of local maxima is correlated with the number of photons.
}
\label{fig:PD}
\end{figure}

The probabilities $\mathcal{P}_n (\alpha, \xi; \pm) = \vert\mathcal{C}_{n}(\alpha,\xi; \pm)\vert^{2}$ of finding $n$ photons in the states $\vert\alpha,\xi; \pm \rangle$ exhibit a shift associated with the distribution of zeros of $P_n (\alpha, \xi; \pm)$. To get some insights on the matter consider the probability distributions of Figure~\ref{fig:PD}. 
The value of $\xi$ has been taken real and is fixed, the figures are plotted in terms of $\vert \alpha \vert$. Observe that $\mathcal{P}_0 (\alpha, \xi; -) =0$, meaning that the classical state $\vert 0 \rangle$ does not contribute to the superposition $\vert \alpha, \xi; - \rangle$. The largest probabilities in the superpositions $\vert \alpha, \xi; \pm \rangle$ are $\mathcal{P}_0 (\alpha, \xi; +)$ and $\mathcal{P}_1 (\alpha, \xi; - )$, corresponding to 0-photon and 1-photon, respectively. The contribution of each photon-number state $\vert n \rangle$ to the squeezing $\vert \alpha, \xi; \pm \rangle$ decreases as $\vert \alpha \vert \rightarrow \infty$, and exhibits $n$ local maxima in $\mathcal{P}_n (\alpha, \xi; +)$ and $\mathcal{P}_{n+1} (\alpha, \xi; - )$. These characteristics show that $\vert\alpha,\xi; + \rangle$ and $\vert\alpha,\xi; -\rangle$ represent strongly different states for $\vert\alpha\vert$ within the convergence radius. The differences between these states are less evident as $\vert \alpha \vert \rightarrow \infty$, which corroborates the results for the trace distance discussed above.

\begin{figure}[htb]
\centering
\subfloat[][$\alpha=0$, $\xi=0.6$]{\includegraphics[width=0.3\textwidth]{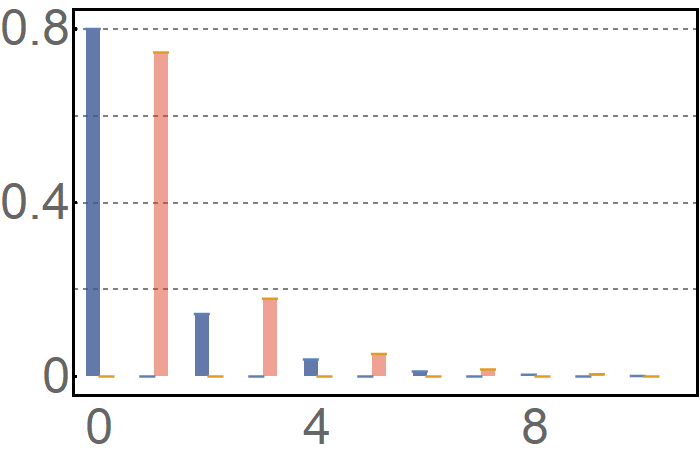}
}
\hskip2ex
\subfloat[][$\alpha=2$, $\xi=0.6$]{\includegraphics[width=0.3\textwidth]{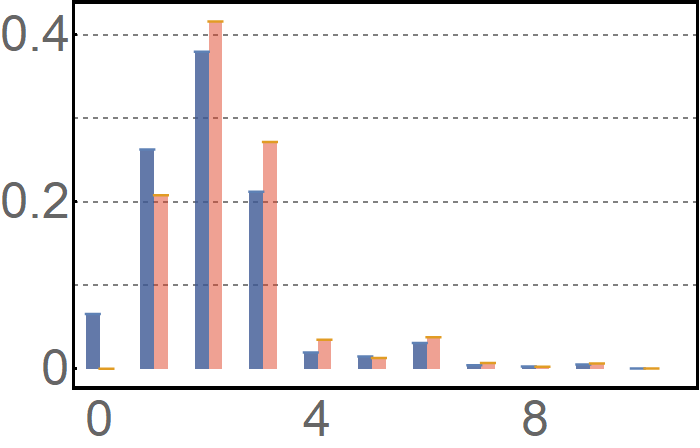}
}
\hskip2ex
\subfloat[][$\alpha=4$, $\xi=0.6$]{\includegraphics[width=0.3\textwidth]{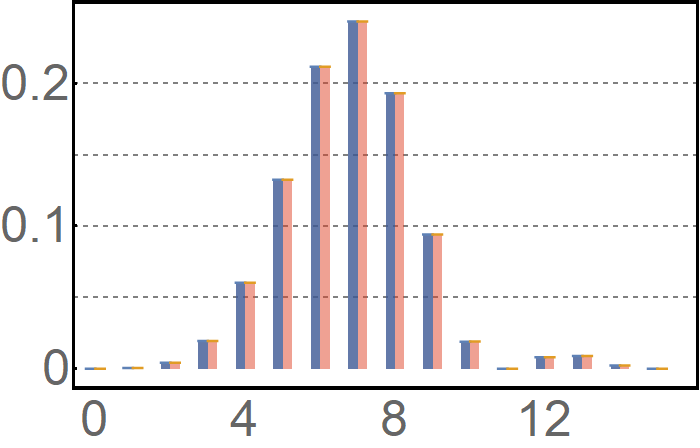}
}
\caption{\footnotesize Histograms counting the presence of $n$-photons in the squeezed states $\vert \alpha, \xi; + \rangle$ (blue) and $\vert \alpha, \xi; - \rangle$ (red) for the indicated values of $\alpha$ and $\xi$. 
}
\label{fig:PD2}
\end{figure}

To get detailed information about the probability distributions shown in Figure~ \ref{fig:PD} we have constructed the histograms exhibited in Figure~\ref{fig:PD2}. There, the probabilities $\mathcal{P}_n (\alpha, \xi; +)$  and $\mathcal{P}_n (\alpha, \xi; -)$ are contrasted for three concrete real values of $\alpha$ and the value of $\xi$ used in Figure~\ref{fig:PD}. In all cases the columns in blue and those in red refer to $\vert \alpha, \xi; + \rangle$ and $\vert \alpha, \xi; - \rangle$, respectively. Figure~\ref{fig:PD2}(a) shows the results for $\alpha=0$. It is highly probable to find zero photons in $\vert 0,\xi; + \rangle$ and one-photon in $\vert 0 , \xi; - \rangle$. The contribution of the remaining probabilities is almost negligible for both vectors. It is then reasonable to consider that $\vert 0 , \xi; - \rangle$ is more nonclassical than $\vert 0 , \xi; + \rangle$. Increasing the value of $\alpha$ in two units the situation changes, see Figure~\ref{fig:PD2}(b). Now the probability of finding zero photons in $\vert 0,\xi; + \rangle$ is very short. The relevant probabilities are concentrated on finding either one, two or three photons, where the probability of finding two photons is the largest one for both states. Nevertheless, $\mathcal{P}_{n} (2,\xi; - ) $ and $\mathcal{P}_{n} (2,\xi; +)$ are still different. One step of two-units further, for $\alpha=4$ in Figure~\ref{fig:PD2}(c), one finds $\mathcal{P}_{n} (4,\xi; +)\simeq \mathcal{P}_{n} (4,\xi; -)$. The latter result is in agreement with the convergence radius evidenced in Figures~\ref{inner} and \ref{fig:PD}.

\subsubsection{Algebraic structure}
\label{algebraic}

The properties of the $P$-functions (\ref{eq:CSSS1}) defining the associated squeezed states $\vert \alpha, \xi; + \rangle$ induce the eigenvalue equation 
\begin{equation}
(a_2 + \xi a_2^{\dagger})\vert\alpha,\xi; -\rangle = \alpha \vert \alpha,\xi; -\rangle,
\label{eq:ALG2}
\end{equation}
with $a_2$ and $a_2^{\dagger}$ a pair of ladder operators arising from supersymmetric quantum mechanics \cite{Mie04}, where they are referred to as {\em distorted} ladder operators \cite{Fer95,Ros96}. The sub-label ``2'' means a particular value of the parameter that characterizes the commutation rules obeyed by these operators: 
\be
[a_2, a_2^{\dagger}] = \left\{
\begin{array}{rl}
0, & \mathcal{H}_0 = \operatorname{span} \{ \vert 0 \rangle \}\\[.6ex]
2, & \mathcal{H}_1 \operatorname{span} \{ \vert 1 \rangle \}\\[.6ex]
1, & \mathcal{H}_s = \operatorname{span} \{ \vert n \rangle, n \geq 2 \}
\end{array}
\right. 
\label{algebra}
\ee
That is, in the distorted representation the Fock space of number states $\mathcal{H}$ is decoupled into the direct sum $\mathcal{H} = \mathcal{H}_0 \oplus \mathcal{H}_1 \oplus \mathcal{H}_s$.

The twice degenerate character of the associated polynomials (\ref{eq:CSSS1}) is therefore expressed through the action of $a_2$ and $a^{\dagger}_2$ on the Fock basis. Namely
\be
a_2 \vert 0 \rangle = a_2 \vert 1 \rangle = 0, \quad a_2^{\dagger} \vert 0 \rangle = 0,
\label{alg1}
\ee
and
\be
a_2 \vert n+1 \rangle = \sqrt{n+1} \vert n \rangle, \quad a_2^{\dagger} \vert n \rangle = \sqrt{n+1} \vert n+1 \rangle, \quad n=1,2, \ldots
\label{alg2}
\ee
Note that both operators $a_2$ and $a_2^{\dagger}$ annihilate the vacuum state $\vert 0 \rangle$, as it is required by the condition $P_0 (\alpha,\xi; -)=0$. 

As the associated squeezed states $\vert \alpha, \xi; - \rangle$ solve the eigenvalue equation (\ref{eq:ALG2}), they minimize the uncertainties associated with the quadratures $x_2 = \frac{1}{\sqrt 2} (a_2^{\dagger} + a_2)$ and $p_2 = \frac{i}{\sqrt 2} (a_2^{\dagger} -a_2)$, which satisfy a commutation rule equivalent to (\ref{algebra}). That is, the set $\vert \alpha, \xi; - \rangle$ is a two-parametric family of minimum uncertainty states for the quadratures $x_2$ and $p_2$.

For $\xi =0$ one has $\vert \alpha, \xi =0 ; - \rangle =\vert \alpha \rangle_{\operatorname{dist}}$, with $\vert \alpha \rangle_{\operatorname{dist}}$ the distorted coherent states introduced in \cite{Fer95,Ros96} and extended to the non-Hermitian case in \cite{Ros18,Zel20}. The relevance of these vectors has been remarked in Section~\ref{asocia}, where we have shown that they define the generating function of the one-photon added states 

\section{Nonclassical properties of the associated squeezed states}
\label{section4}

\begin{figure}[htb]
\centering
\subfloat[][$\alpha=1$]{\includegraphics[width=0.3\textwidth]{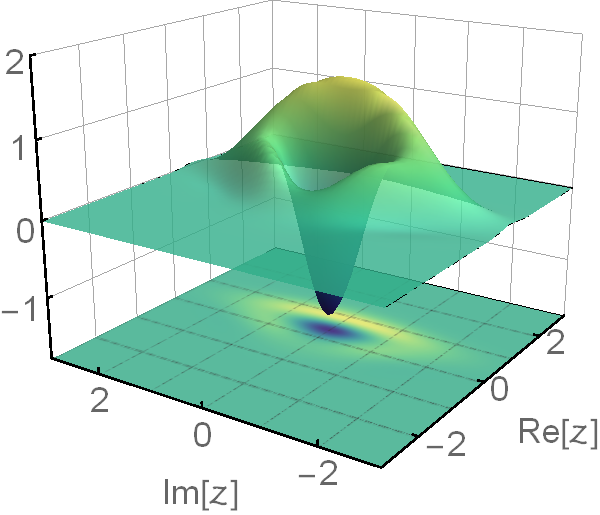}
}
\hskip2ex
\subfloat[][$\alpha=2$]{\includegraphics[width=0.3\textwidth]{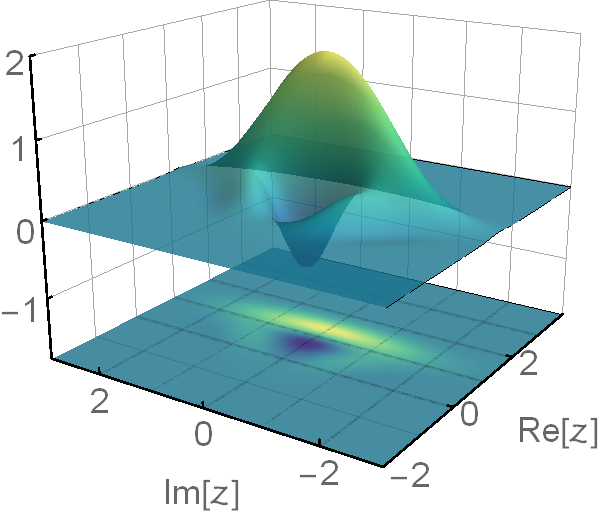}
}
\hskip2ex
\subfloat[][$\alpha=4$]{\includegraphics[width=0.3\textwidth]{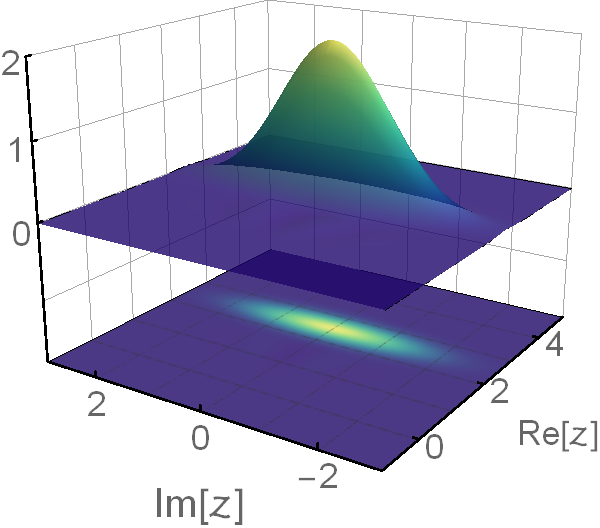}
}
\caption{\footnotesize Wigner function defined by the odd-photon squeezed state $\vert\alpha,\xi; - \rangle$ for $\xi=0.6$ and the indicated values of $\alpha$. The function is plotted on the complex $z$-plane characterizing the Glauber state $\vert z \rangle$ of Eq.~(\ref{eq:NCP2}). 
}
\label{fig:F3}
\end{figure}

By construction, the associated squeezed states $\vert \alpha, \xi; - \rangle$ introduced in Section~\ref{asocia} are also minimum uncertainty states. To analyze their nonclassical properties, consider first the Wigner functions of Figure~\ref{fig:F3}. There, $\alpha$ acquires three different real values while $\xi$ is a fixed real number. These parameters produce the squeezing of the position variable. 

Within the convergence radius discussed in Section~\ref{structure}, the Wigner function of the associated squeezed states $\vert \alpha, \xi; - \rangle$ is markedly different from that of the conventional squeezed states $\vert \alpha, \xi; + \rangle$. In the present case, this difference is illustrated in Figures~\ref{fig:F3}(a) and \ref{fig:F3}(b), where the consistency with the results shown in Figures~\ref{inner}, \ref{fig:PD}, and \ref{fig:PD2} is clear. Outside the radius of convergence, the vectors $\vert \alpha, \xi; \pm \rangle$ tend to represent the same squeezing state since their trace distance $d(\rho_+,\rho_-)$ goes to zero as $\vert \alpha \vert \rightarrow \infty$, see Figure~\ref{inner}. This can be appreciated in Figure~\ref{fig:F3}(c) where the prototypical behavior of the conventional squeezed states is immediately recognized. 

\begin{figure}[htb]
\centering
\subfloat[][$\xi=0$]{\includegraphics[width=0.3\textwidth]{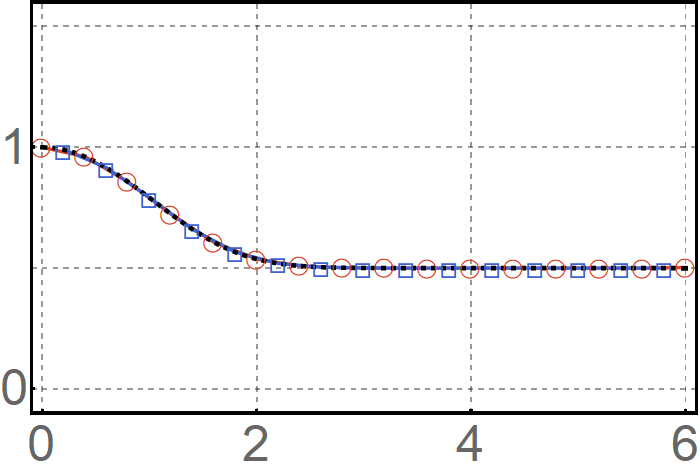}
}
\hskip2ex
\subfloat[][$\xi=0.6$]{\includegraphics[width=0.3\textwidth]{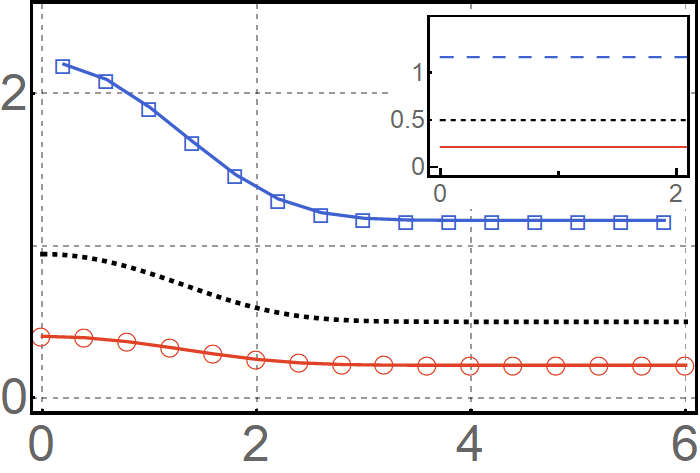}
}
\hskip2ex
\subfloat[][$\xi=0.6$]{\includegraphics[width=0.3\textwidth]{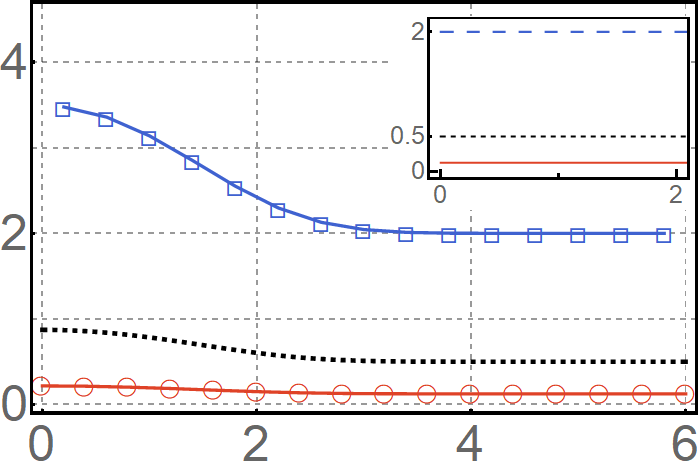}
}
\caption{\footnotesize Variances $\Delta x_2$ (square-blue) and $\Delta p_2$ (circle-red) predicted by the odd-photon squeezed state $\vert \alpha, \xi; - \rangle$ for the indicated values of $\xi$, and as functions of $\vert \alpha \vert$. The inset shows the variances $\Delta x$ (solid-red) and $\Delta p$ (dashed-blue) computed with the conventional squeezed state $\vert \alpha,\xi; + \rangle$. In all cases the minimum uncertainty is shown in dotted-black lines as a reference.
}
\label{fig:F1}
\end{figure}

To go a step further in our analysis we have depicted the behavior of  $\Delta x_2$ and $\Delta p_2$ in Figure~\ref{fig:F1}. These variances have been calculated with the state $\vert \alpha, \xi; - \rangle$ for two real values of $\xi$, as functions of  $\vert \alpha \vert$. For $\xi=0$ both functions coincide, see Figure~\ref{fig:F1}(a), and decay exponentially to the minimum uncertainty value. The latter is evidence of the nonclassical-to-classical transition that is commonly found in the one-photon added coherent states \cite{Aga91}, and verifies our conclusions of Sections~\ref{asocia} and \ref{algebraic} with respect to the relationship between  $\vert \alpha, \xi =0 ; - \rangle= \vert \alpha \rangle_{\operatorname{dist}}$ and $\vert \alpha, 1 \rangle_{\operatorname{add}}$. For $\xi \neq 0$ the variances $\Delta x_2$ and $\Delta p_2$ decay exponentially to the values of $\Delta x$ and $\Delta p $ predicted by the squeezed state $\vert \alpha, \xi; +\rangle$, see Figure~\ref{fig:F1}(b). 

On the other hand, the nonclassical properties of the states $\rho_{\pm}$ can be  also exhibited with the help of a beam splitter \cite{Kim02} (see also \cite{Wan02}). Indeed, if $\rho_{\operatorname{in}}$ represents a bi-partite state of light entering a 50:50 beam splitter, the off-diagonal elements of the output state $\rho_{\operatorname{out}}$ certify nonclassical correlations if they are different from zero. In such a case, measuring the number of photons at one of the output ports of the beam splitter is affected by the result of detecting photons at the other port and vice versa \cite{Zel17a}. Therefore, the nontrivial off-diagonal elements of $\rho_{\operatorname{out}}$ reveal the nonclassicality of $\rho_{\operatorname{in}}$, at least in one of its two channels \cite{Kim02}

A measure of the above notion of nonclassicality is provided by the linear entropy $S (\rho) =1- \operatorname{Tr} \rho^2$ \cite{Nie00}, which quantifies the purity of any quantum state $\rho$. In general $0 \leq S \leq \frac{D-1}{D}$, with $D \geq 2$ the order of the square matrix representing the state $\rho$. In the present case, $0 \leq S \leq 1$, with $S=0$ for $\rho$ a pure state and $S=1$ for $\rho$ a completely mixed state. The technique exploits the fact that a completely entangled pure state $\rho_{\operatorname{out}}$ is such that its reduced states are completely mixed. Therefore, if $\rho_{\pm}$ is in channel 1 of the bi-partite state $\rho_{\operatorname{in}}$, the linear entropy of the reduced output state $\widetilde \rho_{\pm} =\operatorname{Tr}_2 \rho_{\operatorname{out}}$ will provide a measurement of the nonclassicality of $\rho_{\pm}$, which is ranked from 0 to 1.

\begin{figure}[htb]
\centering
\includegraphics[width=0.4\textwidth]{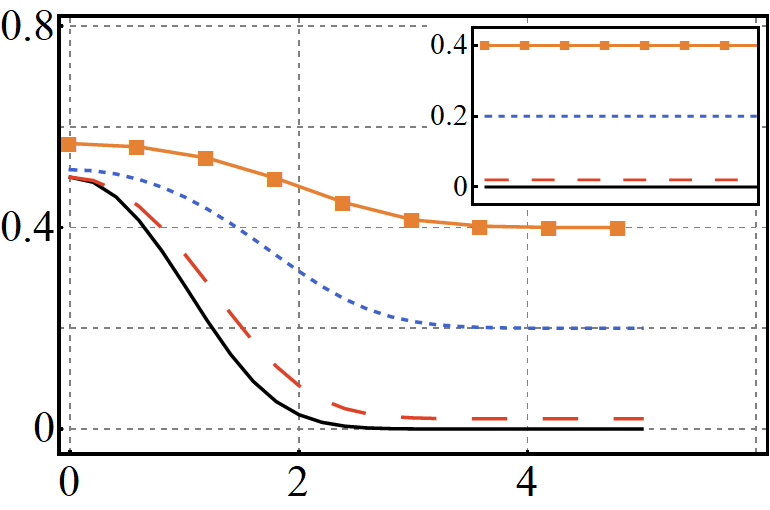}

\caption{\footnotesize
Linear entropy of the reduced state $\widetilde \rho_-$ corresponding to the odd-photon squeezed state $\vert \alpha,\xi; - \rangle$ at the output of a 50:50 beam-splitter. The horizontal axis refers to $\vert\alpha\vert$,  the plots correspond to $\xi=0$ (solid-black), $\xi=0.2$ (dashed-red), $\xi=0.6$ (dotted-blue), and $\xi=0.8$ (square-orange). The inset corresponds to the linear entropy of the conventional squeezed state $\vert \alpha,\xi; +\rangle$.
}
\label{fig:F5}
\end{figure}

Figure~\ref{fig:F5} shows the linear entropy $S(\widetilde \rho_-)$ associated with the odd-photon squeezed state $\vert \alpha,\xi; - \rangle$ for different values of $\xi \in \mathbb R$, and as a function of $\vert \alpha \vert$. In all cases the purity of $\widetilde \rho_-$ is larger at $\vert \alpha \vert =0$. As in the previous measures of nonclassicality, the state $\vert \alpha,\xi; - \rangle$ becomes more classical as $\vert \alpha \vert$ increases. Outside the radius of convergence, the quantumness of the odd-photon squeezed state is bounded from below by the nonclassicality of the conventional squeezed states $\vert \alpha,\xi; + \rangle$.

\section{Concluding remarks}
\label{conclu}

We have shown that carrying out a thorough search of the solutions of the recurrence relations gives rise to unnoticed squeezed states of light. 

The constraint $P_{1} =\alpha P_{0}$ included in Eq.~(\ref{eq:NSS5-x}), additional to the three-term recurrence relation (\ref{eq:NSS5}), results very restrictive to define minimum uncertainty states. The latter because the initial value $P_0 \neq 0$ is then imposed by necessity in the conventional searching of squeezed states. Such a constraint has kept hidden the possibility of finding other solutions to the problem of getting squeezed states of light. 

We have overpassed the initial value $P_0 \neq 0$ by solving first, in general form, the recurrence relation (\ref{eq:NSS5}) without imposing any initial value a priori. We faced the problem within the theory of finite differences and orthogonal polynomials. Once the general solution is achieved, a second step consists in determining the initial conditions that each of the two independent solutions must satisfy in order to be well defined. 

The conventional minimum uncertainty states that lead to the squeezing of either of the field quadratures have been recovered as particular cases.

The method presented in this work also produces superpositions of photon-number states whose coefficients are determined by the associated Hermite polynomials. As far as we know, this class of squeezed states has not been reported anywhere before the present work.

\section*{Acknowledgments}

This research has been funded by Consejo Nacional de Ciencia y Tecnolog\'ia (CONACyT), Mexico, grant number A1-S-24569. V.~Hussin acknowledges the research grant received from the Natural Sciences and Engineering Research Council (NSERC) of Canada. K.~Zelaya acknowledges the support from the Fonds de recherche du Qu\'ebec--Nature et technologies (FRQNT), international internship award 210974.

K.~Zelaya would like to thank Professor Veronique Hussin and the Centre de Recherches Math\'ematiques for their kind hospitality.

\appendix
\section{Solving recurrence relations by the comparison method}
\label{ApA}
\setcounter{section}{0}  

\renewcommand{\thesection}{A}

\renewcommand{\theequation}{A-\arabic{equation}}
\setcounter{equation}{0}  

In this appendix we construct the general solution of recurrence relations before considering the initial conditions. Our approach, hereafter referred to as {\em comparison method}, is addressed to determine whether the solutions of two recurrence relations can be paired by assuming that one of the recurrence problems is already solved. We focus on three-term recurrence relations but the procedure is easily adapted to recurrences having a different number of terms.

Consider the recurrences
\begin{equation}
a_{n} f_{n+1}+ b_{n}f_{n}+ d_{n}f_{n-1}=0, \quad a_n, b_n, d_n\in\mathbb{C}, \quad n=1,2, \ldots \, ,
\label{eq:TTRR1}
\end{equation}
and
\begin{equation}
A_{n}  F_{n+1} + B_{n} F_{n} + D_{n} F_{n-1}=0, \quad A_n, B_n, D_n\in\mathbb{C}, \quad n=1,2, \ldots
\label{eq:TTRR2}
\end{equation}
Assuming that the set $F_n$ defined by (\ref{eq:TTRR2}) is already known, we want to determine whether $f_{n}$ can be written in terms of  $F_{n}$. The affirmative answer depends strictly on the profile and properties of $F_{n}$. 

Let $h_n$ be a function such that $f_n= h_n F_n$. Introducing it into Eq.~\eqref{eq:TTRR1} and comparing the result with \eqref{eq:TTRR2} we arrive at the relationships
\begin{equation}
\frac{h_{n+1}}{h_n}=\frac{A_n b_n}{B_n a_{n}}, \qquad \frac{h_{n+1}}{h_{n}}=\frac{B_{n+1} d_{n+1}}{D_{n+1} b_{n+1}}.
\label{eq:TTRR3}
\end{equation}
Both equations in (\ref{eq:TTRR3}) should lead to the same function $h_n$, so we impose the compatibility condition
\begin{equation}
\frac{A_{n} D_{n+1} b_{n}b_{n+1}}{B_{n} B_{n+1} a_{n}d_{n+1}}=1, \quad n=0,1,2, \ldots
\label{eq:TTRR4}
\end{equation}
If (\ref{eq:TTRR4}) is fulfilled, the auxiliary function is obtained by solving any of the recurrence relations in~\eqref{eq:TTRR3}. One finds
\begin{equation}
h_{n+1}=h_0 \prod_{k=0}^{n}\frac{A_k b_k}{B_k a_k}= 
h_0 \prod_{k=0}^{n}\frac{B_{k+1} d_{k+1}}{D_{k+1} b_{k+1}},  \quad n=0,1, \ldots,
\label{eq:TTRR5}
\end{equation}
where $h_0$ may be determined from the initial conditions trhough $f_{n}=h_{n} F_{n}$. 

\subsection{The three-term recurrence relation of Section~\ref{3term}}

We apply the comparison method to solve the recurrence relation (\ref{eq:NSS5}) of  Section~\ref{3term}. First we show the way in which the conventional results are recovered and then we obtain more general results.

\subsubsection{Usual solution}

Let us rewrite the  recurrence relation (\ref{eq:NSS5}) as follows
\begin{equation}
P_{n+1}- \alpha P_{n}+  \xi n P_{n-1}=0, \quad n=1,2, \ldots
\label{eq:TTRR5-1}
\end{equation}
To apply the comparison method we use $f_{n}=P_{n}(\alpha,\xi)$, with $a_{n}=1$, $b_n=\alpha$ and $d_{n}= \xi n$. 

Exploring the well known recurrence relations for the classical orthogonal polynomials \cite{Gra07} we find that the recurrence relation for the Hermite polynomials
\begin{equation}
H_{n+1}(z)-2zH_{n}(z)+2nH_{n-1}(z)=0 \, ,\quad n=1,2, \ldots,
\label{eq:TTRR5-2}
\end{equation}
is useful in the present case. That is, taking $F_{n}=H_{n}(z)$, with $A_{n}=1$, $B_{n}=-2z$, and $D_{n}=2n$, the compatibility condition~\eqref{eq:TTRR5} is fulfilled with $z=\alpha/\sqrt{2\xi}$. Therefore, we obtain $h_{n}=h_0 \left(\frac{\xi}{2}\right)^{n/2}$. 

Now, taking into account the constraint (\ref{eq:NSS5-x}), meaning $P_0 \neq 0$, we may fix $h_0$ by the initial condition $P_{0}(\alpha,\xi)=1$. Therefore, $f_{n} = h_{n} F_{n}$ yields the well known result $P_{n}(\alpha,\xi)=(\xi/2)^{n/2}H_{n}(\alpha/\sqrt{2\xi})$. These roots of the recurrence problem (\ref{eq:NSS5})-(\ref{eq:NSS5-x}) have been used to recover the expression of the conventional squeezed states $\vert \alpha, \xi; + \rangle$ in Eq.~(\ref{eq:NSS6}) of the main text.

\subsubsection{General solution}

Looking for a general solution one should recall that the confluent hypergeometric function ${}_{1}F_{1}(a,c;z) \equiv M(a,c; z)$, with $a =-n$ and $c \neq -m$ yields a polynomial of degree $n$ in $z$ \cite{Olv10} ($n$ and $m$ positive integers). Then, we may wonder whether the solutions of (\ref{eq:TTRR5-1}) can be paired with such polynomials. A first insight is obtained by comparing the confluent hypergeometric recurrence relation 
\be
(c-a) M(a-1,c;z) + (2a-c+ z) M(a,c;z) -a M(a+1,c;z)=0, \quad a=-n, c\neq -m,
\label{eme}
\ee
with Eq.~(\ref{eq:TTRR5-1}) since it makes clear that the compatibility condition (\ref{eq:TTRR4}) cannot be achieved. Nevertheless, decoupling (\ref{eq:TTRR5-1})  into even and odd values of $n$ we respectively have
\begin{equation}
P_{2n+2} (\alpha,\xi) + \left( 4 \xi  n + \xi-\alpha^2 \right) P_{2n} (\alpha,\xi)+ 4 \xi^2  n \left( n-\tfrac12 \right) P_{2n-2} ( \alpha, \xi )=0,
\label{eq:TTRR7}
\end{equation}
and
\begin{equation}
P_{2n+3} (\alpha,\xi) +  \left( 4 \xi n + 3\xi-\alpha^2 \right) P_{2n+1} (\alpha, \xi)+ 4 \xi^2 n \left( n + \tfrac12 \right) P_{2n-1}(\alpha,\xi)=0.
\label{eq:TTRR9}
\end{equation}
These results are now compatible with (\ref{eme}) for either $c=\tfrac12$ or $c= \tfrac32$. It is useful to recall the relationship between the confluent hypergeometric function and the Hermite polynomials
\be
M(-n, \tfrac12; x^2)= (-1)^{n} \frac{n!}{(2n)!} H_{2n}(x), \quad M(-n, \tfrac32; x^2)= \frac{ (-1)^{n} }{2x} \frac{n!}{(2n+1)!} H_{2n+1}(x).
\ee
Thus, in the present case we may consider $F_n = M(-n,c;z)$ with $c$ equal to either $1/2$ or $3/2$ in order to get the corresponding auxiliary function (\ref{eq:TTRR5}). The latter provides a first solution to the problem. A second solution can be obtained by recalling that the confluent hypergeometric equation admits two linearly independent solutions. Given $y_1=M(a,c;z)$, the function $y_4=z^{1-c} e^z M(1-a, 2-c, -z)$ is such that $W(y_1, y_4) =(1-c) z^{-c} e^z$ \cite{Olv10}, so that $y_1$ and $y_2$ are linearly independent if $c\neq 1$. Therefore, if $f_{2n} = h_{2n} M(-n, \tfrac12, z)$ is our first solution, we may write $\widetilde h_{2n} M(1+n, \tfrac32, -z)$ for the second one, with $\widetilde h_{2n}$ absorbing the factors $z^{1/2} e^z$ and being to be determined. In this form, the general solution for the even labels $2n$ is written as a linear combination of the above functions. The straightforward calculation yields
\begin{equation}
P_{2n} = (-2\xi)^{n} \left[ \left( \tfrac12 \right)_{n} \kappa_{1}\,  M \!\left(-n,\tfrac12; \tfrac{\alpha^{2}}{2\xi}\right) + n! \, \widetilde{\kappa}_{1} \, M\!
\left( n+1, \tfrac{3}{2};- \tfrac{\alpha^{2}}{2\xi}\right) \right],
\label{eq:TTRR8}
\end{equation}
where the complex-valued coefficients $\kappa_{1} (\alpha,\xi)$ and $\widetilde{\kappa}_{1} (\alpha,\xi)$ are fixed by the initial conditions. Equivalently, for the odd labels $2n+1$, we have
\begin{equation}
P_{2n+1} = (-2\xi)^{n} \left[\left( \tfrac32 \right)_n \kappa_{2} M \! \left(-n,\tfrac{3}{2};\tfrac{\alpha^{2}}{2\xi}\right) + n! \, \widetilde{\kappa}_{2}  M\! \left(n+1,\tfrac{1}{2};-\tfrac{\alpha^{2}}{2\xi}\right) \right],
\label{eq:TTRR10}
\end{equation}
with $\kappa_{2} (\alpha,\xi)$ and $\widetilde{\kappa}_{2} (\alpha,\xi)$ defined by the initial conditions. 

$\bullet$ {\bf Solutions obeying the constraint (\ref{eq:NSS5-x}).}

\noindent
Taking into account the constraint (\ref{eq:NSS5-x}), that is $P_0 \neq 0$, we may take $P_{0}(\alpha,\xi)=1$. Then $P_{1}(\alpha,\xi)=\alpha$, and
\[
\widetilde\kappa_{1} (\alpha,\xi)= \widetilde{\kappa}_{2} (\alpha,\xi)=0, \quad \kappa_{1} (\alpha,\xi)= 1, \quad {\kappa}_{2} (\alpha,\xi)= \alpha.
\]
The above results permit to recover the well known expression of the squeezed states \eqref{eq:NSS6}.

$\bullet$ {\bf Solutions that do not satisfy the constraint (\ref{eq:NSS5-x}).}

\noindent
Making $P^{(2)}_{0}(\alpha,\xi)=0$ and $P^{(2)}_{1}(\alpha,\xi)=1$ we find $\widetilde\kappa_{1} (\alpha,\xi)= \widetilde{\kappa}_{2} (\alpha,\xi)=1$, and
\[
\kappa_{1} (\alpha,\xi)= -M \! \left(1, \tfrac32;- \tfrac{\alpha^{2}}{2\xi} \right), \quad {\kappa}_{2} (\alpha,\xi)= \tfrac{\alpha^2}{\xi} M \! \left(1, \tfrac32;- \tfrac{\alpha^{2}}{2\xi} \right).
\]
After some calculations, from the above expressions one arrives at the results presented in Eqs.~\eqref{eq:CSSS1-2} and \eqref{eq:CSSS1-3} of the main text.

\appendix
\setcounter{section}{1}  
\section{Orthogonal and associated polynomials}
\label{ApB}

\renewcommand{\thesection}{B-\arabic{section}}

\renewcommand{\theequation}{B-\arabic{equation}}
\setcounter{equation}{0}  

Following \cite{Ass91} we consider a set $\{ p_n(x) \}$ of orthogonal polynomials
\[
p_n (x) = \gamma_n x^n + \gamma_{n-1} x^{n-1} + \cdots + \gamma_0, \quad \gamma_n >0,
\]
that satisfy the three-term recurrence relation
\be
x p_n(x) = a_{n+1} p_{n+1} (x)  + b_n p_n(x) + a_n p_{n-1}(x), \quad n \geq 0,
\label{Fav1}
\ee
with initial values
\be
 p_{-1}(x) =0, \quad p_0(x) =1,
\label{Fav2}
\ee
and
\be
a_n = \frac{\gamma_{n-1}}{\gamma_n} >0, \quad b_n = \int x p_n^2(x) d\mu(x) \in \mathbb R.
\label{Fav3}
\ee
The function $\mu(x)$ in the recurrence coefficients (\ref{Fav3}) is a probability measure on the real line such that
\begin{equation}
\int p_{n}(x)p_{m}(x) d\mu(x)=\delta_{nm}, \quad m, n \geq 0.
\label{eq:OAP1}
\end{equation}
Markedly, except for the classical orthogonal polynomials \cite{Olv10}, finding the measure $\mu$ and the solutions $p_n(x)$ of the system (\ref{Fav1})-(\ref{eq:OAP1}) represents a formidable amount of work in general. In this respect the Favard's theorem \cite{Fav35} (see also \cite{Chi78}) is very useful since it states that for the recurrence problem defined by (\ref{Fav1})-(\ref{Fav2}),  there exists a probability measure $\mu$ so that the recurrence coefficients acquire the form (\ref{Fav3}) and the orthogonality (\ref{eq:OAP1}) is satisfied \cite{Ass91}, and vice versa. Therefore, it is natural to concentrate in solving (\ref{Fav1})-(\ref{Fav2}) and then to allude the Favard's theorem to ensure orthogonality.

A slight alteration of the three-term recurrence relation (\ref{Fav1}) produces new results. Namely, 
\be
x p_n^{(k)} (x) = a_{n+k+1} p_{n+1}^{(k)} (x)  + b_{n+k}  p_n^{(k)} (x) + a_{n+k}  p_{n-1}^{(k)} (x), \quad n \geq 0,
\label{Fav4}
\ee
with
\be
 p_{-1}^{(k)} (x) =0, \quad p_0^{(k)} (x) =1,
\label{Fav5}
\ee
defines the $k$th associated orthogonal polynomials $p_n^{(k)}(x)$ \cite{Ass91,Bel90} (associated to the ones with $k=0$), also called numerator polynomials \cite{Chi78}. Given $k$, the set $\{ p_n^{(k)}(x) \}$ defines a solution of the recurrence relation (\ref{Fav1}) with $\mu^{(k)}$ the corresponding measure (guidelines for determining $\mu$ can be found in \cite{Chi78}, \cite{Rah01}, and references quoted therein). The associated recurrence problem (\ref{Fav4})-(\ref{Fav5}) is very useful for the purposes of this work since it permits to avoid the strong restricttion $P_0 \neq 0$ from the constraint (\ref{eq:NSS5-x}).

In the main text we work with (\ref{Fav1}) rewritten in the form \cite{Chi78,Sze59}
\begin{equation}
p_{n+1}=(x-c_{n})p_{n}-\lambda_{n}p_{n-1},  \quad p_{-1}=0,  \quad p_{0}=1, \quad n=0,1,2,\ldots,
\label{eq:OAP2}
\end{equation}
where the coefficients $c_n$ and $\lambda_n$ are complex in general. 

We identify (\ref{eq:OAP2}) with a second order difference equation \cite{Mil33}, so there are two independent solutions for each value of $n$. If $p_{n}$ and $g_{n}$ solve a difference equation, they are independent if
\begin{equation}
\mathcal{C}(p_n,g_n)= \operatorname{Det}\left( \begin{alignedat}{3} \,&p_{n} \quad && g_{n} \\ &p_{n+1} \quad && g_{n+1} \end{alignedat}\right) \neq 0.
\label{eq:OAP3}
\end{equation}
The above determinant is known as the Casorati function \cite{Mil33}, and is the discrete version of the Wronskian in differential equations.

Given a first solution $p_{n}$, a second solution $g_{n}$ such that $\mathcal{C}(p_n,g_n) \neq 0$ may be constructed by reducing the order of the corresponding difference equation \cite{Jer96}. The method is summarized with the algorithm
\begin{equation}
\frac{g_{n+1}}{p_{n+1}}=d_{0} + d_1 \sum_{m=0}^{n} \frac{\mathcal{R}(m)}{p_{m}p_{m+1}} \, , \quad g_{0}=0 \, , \quad \mathcal{R}(m)=\prod_{k=0}^{m-1}\lambda_k \, ,
\label{eq:OAP4}
\end{equation}
with $d_{0}$ and $d_{1}$ arbitrary complex constants. Nevertheless, such a method yields unnecessary complications. We circumvent them by considering the additional difference problem
\begin{equation}
p_{n+1}^{(1)}=(x-c_{n+1}) p_{n}^{(1)} -\lambda_{n+1}p_{n-1}^{(1)}, \quad p_{-1}^{(1)}=0 \, , \quad p_{0}^{(1)}=1, \quad n=0,1, \ldots,
\label{eq:OAP5}
\end{equation}
where $\lambda_{n}$ and $c_{n}$ are the same as those defining \eqref{eq:OAP2}. Thus, we pay attention to the associated polynomials $p_{n}^{(1)} (x)$ of the $p_n(x)$ that solve (\ref{eq:OAP2}).

Making $g_{n}=p_{n-1}^{(1)}$, and considering the initial condition $p_{-1}^{(1)}=0$, produces $g_{0}=0$, which is the initial condition considered in the algorithm \eqref{eq:OAP4}. Therefore, the difference problem \eqref{eq:OAP5} acquires the form
\begin{equation}
g_{n+1}=(x-c_n)g_{n}-\lambda_n g_{n-1}, \quad  g_{0}=0, \quad g_{1}=1.
\label{eq:OAP7}
\end{equation}
It is now clear that the associated polynomials $g_n$ do not satisfy the constraint (\ref{eq:NSS5-x}) since $g_0 =0$. Besides, the Casorati function between $p_{n}$ and $g_{n}$ is $C(p_n,g_n)=\lambda_1$~\cite{Ass91}, so these solutions are independent, provided that $\lambda_1 \neq 0$. That is, the second independent solution $g_{n}$ given by Eq.~\eqref{eq:OAP4} may be also computed from the recurrence relation \eqref{eq:OAP7}. 


\end{document}